\documentclass{aa}  
\usepackage{graphicx}
\usepackage[varg]{txfonts}
\usepackage{float}
\usepackage{placeins,afterpage}
\usepackage[caption = false]{subfig}
\usepackage[colorlinks,citecolor=blue]{hyperref}
\hypersetup{
    colorlinks = true,
    linkcolor = blue,
    anchorcolor = blue,
    citecolor = blue,
    filecolor = blue,
    urlcolor = blue
    }

\begin{document} 

    \title{The EDIBLES survey}   
    \subtitle{X. The 6196\,{\AA}~diffuse interstellar band:\\Identification of side DIBs as an indication of a small carrier molecule}
    
\author{
    A.~Ebenbichler\inst{1}
    \and
    M.~Ončák\inst{2}
    \and
    N.~Przybilla\inst{1}
    \and
    H.~R.~Hrodmarsson\inst{3}
    \and
    J.~V.~Smoker\inst{4,5}
    \and
    R.~Lallement\inst{6}
    \and
    A.~Farhang\inst{7}
    \and 
    C.~Bhatt\inst{8,9}
    \and
    J.~Cami\inst{8,9,10}
    \and
    M.~Cordiner\inst{11,12}
    \and 
    P.~Ehrenfreund\inst{13}
    \and
    N.~L.~J.~Cox\inst{14}
    \and
    J.~Th.~van~Loon\inst{15}
    \and
    B.~Foing\inst{13,16,17}
}

    \institute{Universit\"at Innsbruck, Institut f\"ur Astro- und Teilchenphysik, Technikerstr. 25/8, 6020 Innsbruck, Austria\\
        \email{Alexander.Ebenbichler@uibk.ac.at}
        \and
        Universit\"at Innsbruck, Institut f\"ur Ionenphysik und Angewandte Physik, 
        Technikerstr. 25, 6020 Innsbruck, Austria
        \and
         Univ Paris Est Creteil and Université Paris Cité, CNRS, LISA, F-75013 Paris, France
        \and
        European Southern Observatory, Alonso de Cordova 3107, Vitacura, Santiago, Chile%% Jonathan Smoker 1
        \and  
        UK Astronomy Technology Centre, Royal Observatory, Blackford Hill, Edinburgh EH9 3HJ, UK%% Jonathan Smoker 2
        \and
        GEPI, Observatoire de Paris, PSL Research University, CNRS, Universit\'e Paris-Diderot, Sorbonne Paris Cit\'e, Place Jules Janssen, 92195 Meudon, France%% Rosine Lallement
        \and
        School of Astronomy, Institute for Research in Fundamental Sciences, 19395-5531 Tehran, Iran%% Amin Farhang
        \and
        Department of Physics and Astronomy, The University of Western Ontario, London, ON N6A 3K7, Canada%% Jan Cami 1
        \and
        Institute for Earth and Space Exploration, The University of Western Ontario, London, ON N6A 3K7, Canada%% Jan Cami 2
        \and
        SETI Institute, 189 Bernardo Ave, Suite 100, Mountain View, CA 94043, USA%% Jan Cami 3
        \and
        Astrochemistry Laboratory, NASA Goddard Space Flight Center, Code 691, 8800 Greenbelt Road, Greenbelt, MD 20771, USA %% Cordiner
        \and
        Department of Physics, The Catholic University of America, Washington, DC 20064, US %% Martin Cordiner
        \and
        Laboratory for Astrophysics, Leiden Observatory, Leiden University, 
        P.O.~Box 9513, 2300 RA Leiden, The Netherlands%% Harold Linnartz, Pascale Ehrenfreund
        \and
        Centre d’Etudes et de Recherche de Grasse, ACRI-ST, Av. Nicolas Copernic, Grasse 06130, France%% Nick Cox
        \and
        Lennard-Jones Laboratories, Keele University, ST5 5BG, UK%% Jacco v. Loon
        \and
        ESTEC, ESA, Keplerlaan 1, 2201 AZ Noordwijk, The Netherlands
        \and
        LUNEX EMMESI Euromoonmars Earth Space Innovation, SBIC Noordwijk, The Netherlands
         }

   \date{Received 2 December 2024; accepted 5 February 2025}

\abstract
% context heading (optional)
{Numerous studies of diffuse interstellar band (DIB) profiles have detected  substructures, which in turn suggests that large molecules are acting as their carriers.
However, some of the narrowest DIBs generally do not show such substructures, suggesting the possibility of very small carriers.}
% aims heading (mandatory)
{Based on the previously found tight correlation of the three narrow DIBs at 6196, 6440, and 6623\,{\AA} and the present detection of weaker side DIBs to each of them in the extensive  dataset from the ESO Diffuse Interstellar Bands Large Exploration Survey, we investigated whether they may stem from small linear carrier molecules.
This approach can lead to concrete DIB carrier suggestions, which can be tested in laboratory measurements in future studies.}
% methods heading (mandatory)
{We suggest that the DIBs we studied here represent individual rotational transitions of a small molecule. We determined the molecular constants from observations and compared them with data from a large set of quantum-chemical calculations to constrain possible carrier candidates. Furthermore, we determined the rotational temperatures by fitting line ratios using the fitted molecular models.}
% results heading (mandatory)
{We determined molecular constants for three DIB systems and the corresponding transition types. The fitted rotational temperatures lie within the range of known interstellar diatomic molecules. We identified several DIB carrier candidates, almost all of them molecular ions. 
%through quantum-chemical calculations.
Some of them are metastable species, indicating the possibility of collision complexes as DIB carriers.}
% conclusions heading (optional), leave it empty if necessary 
{If our hypothesis holds, this would be a major step  towards the identification of a carrier molecule of the 6196\,{\AA} DIB, the strongest among the narrow DIBs.}


\maketitle
%
%-------------------------------------------------------------------

\section{Introduction}
Identifying the  carriers of the diffuse interstellar bands (DIBs) has been a challenge of astrophysics for many years, since their first discovery more than a century ago \citep{Heger22}. To date, the number of known DIBs has increased to more than 600, covering spectral features ranging in wavelength from the optical blue \citep[e.g.][]{Hobbsetal09} to the near-infrared \citep[near-IR; e.g.][] {Fanetal22,ebenbichler22,2022hamano}. Despite considerable efforts in laboratory spectroscopy, their carriers remain elusive, with the exception of C$_{60}^+$, which gives rise to four DIBs in the near-IR \citep[e.g.][]{2015Natur.523..322C,2017ApJ...846..168S,2019ApJ...875L..28C,2020JMoSp.36711243L}. Consequently, more than 99\% of the DIBs still lack an identified carrier molecule.

\begin{figure}[ht]
\centering
\subfloat[]{
\includegraphics[width=.97\hsize,trim={120 20 120 20          },clip]{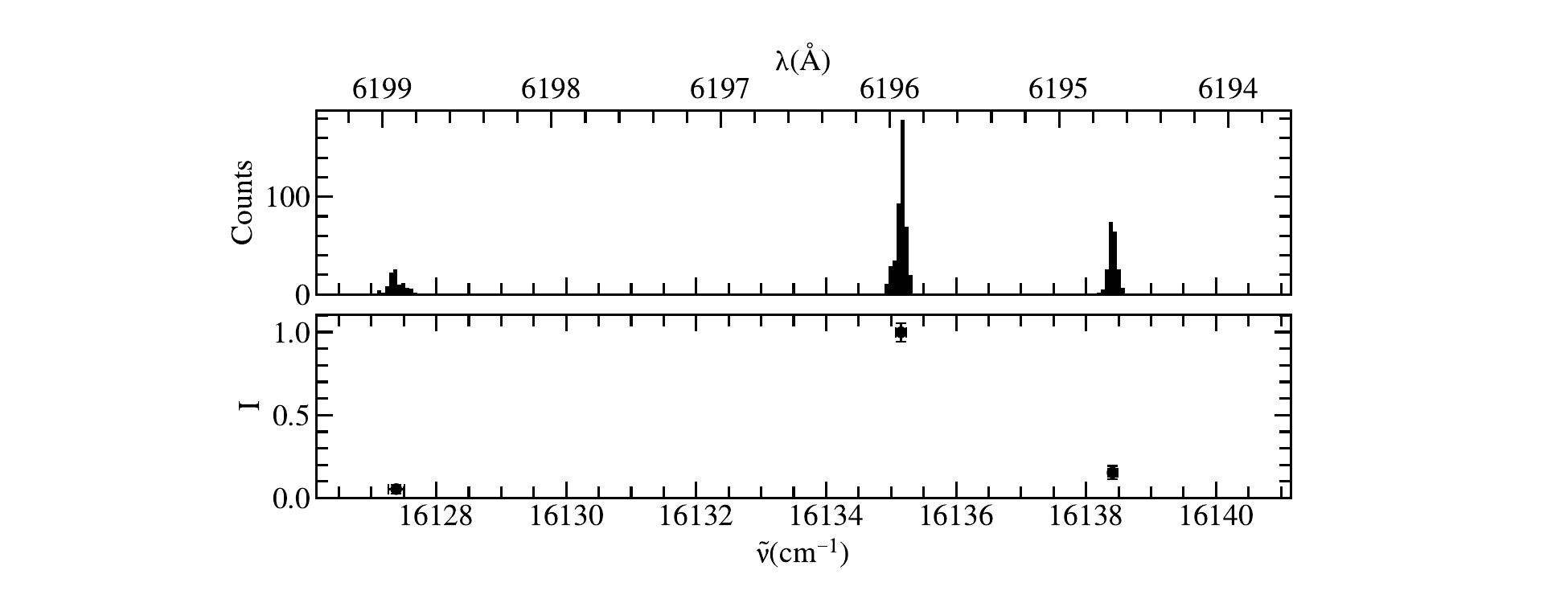}
     \label{fig:hist_and_strength_6196}
}\\
\subfloat[]{
\includegraphics[width=.97\hsize,trim={120 20 120 20        },clip]{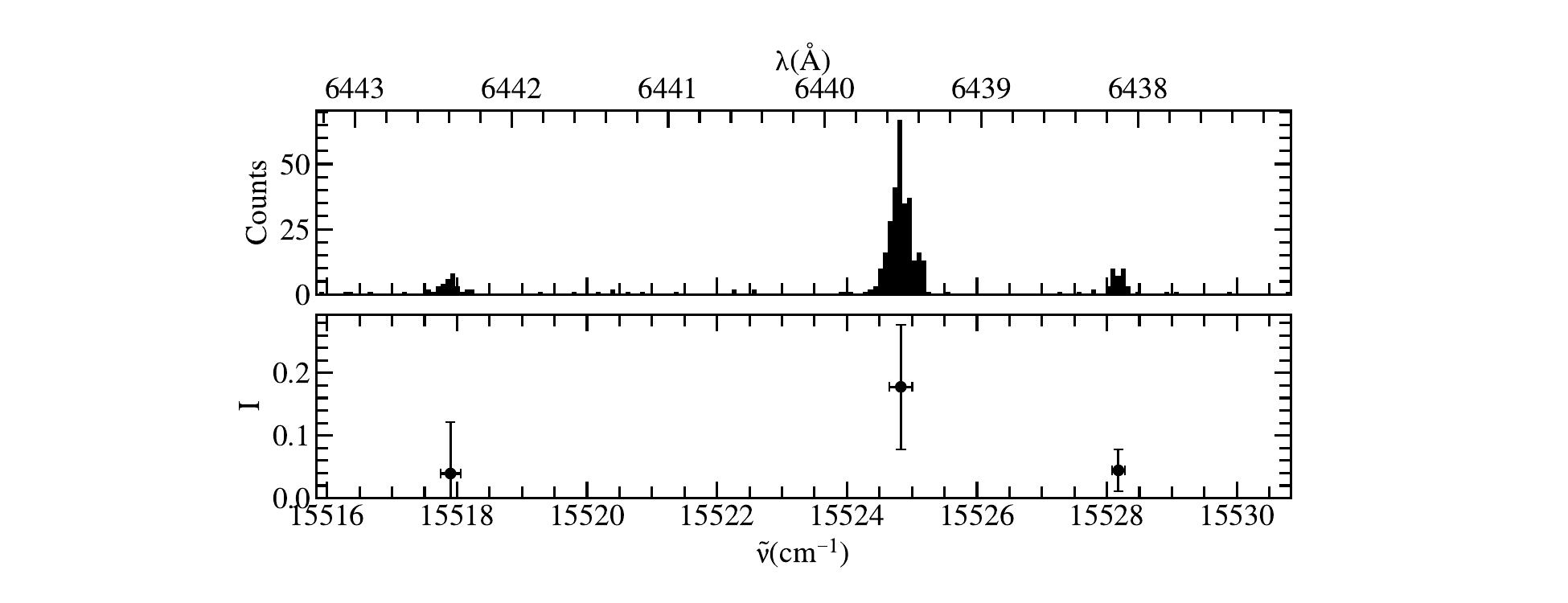}
     \label{fig:hist_and_strength_6440}
}\\
\subfloat[]{
\includegraphics[width=.97\hsize,trim={120 20 120 20      },clip]{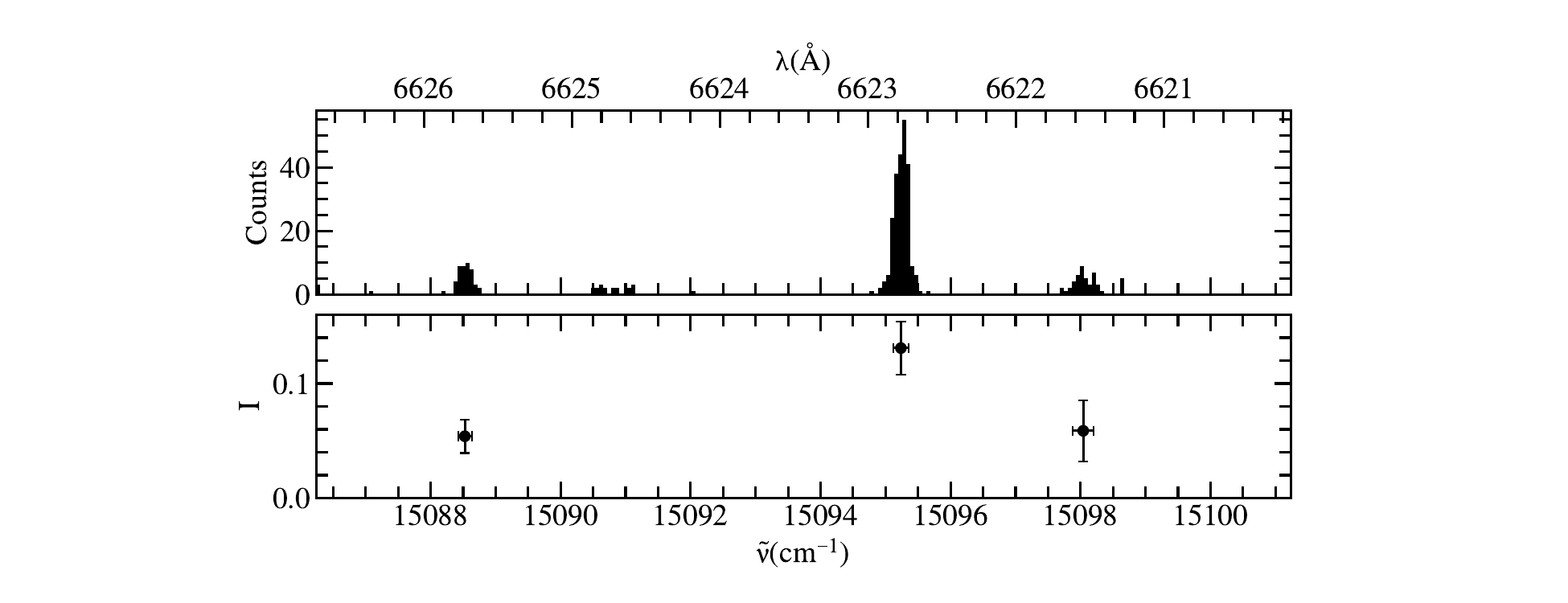}
     \label{fig:hist_and_strength_6623}
}
  \caption{Histogram (top) and median strength ratios (bottom) of the proposed side band matches of the 6196 (panel~\ref{fig:hist_and_strength_6196}), 6440 (\ref{fig:hist_and_strength_6440}), and 6623\,{\AA}~DIBs (\ref{fig:hist_and_strength_6623}) from DIB alignment, using the full EDIBLES sample.
  The strength ratios are relative to the 6196\,{\AA}~DIB.}
\label{fig:hist_and_strength}
\end{figure}

However, numerous laboratory, theoretical, and observational investigations have yielded insights into potential DIB carriers. Due to their wavelength stability, it is believed that DIBs are caused by molecules in the gas phase \citep{1995herbig}. Observations indicate that the nature of these molecules can be of a wide variety. 
Some very broad DIBs show a Lorentzian profile shape, for example the 4430 \citep{2002ApJ...567..407S} and 5450\,{\AA}~DIB. In the case of the 5450\,{\AA}~DIB, an acetylene (C$_{2}$H$_{2}$) plasma absorption spectrum coincided in wavelength, band width, and band shape \citep{linnartz2010}. Unfortunately, the carrier that formed in the plasma could not be unambiguously identified in that work due to the short lifetime of the transition of $\tau$\,$\approx$\,0.15\,ps, but it certainly contains hydrogen and carbon. 
Many classes of molecules have been investigated as possible carriers, such as carbon chain radicals \citep{2000motylewski}, cationic carbon rings of various lengths \citep{2022rademacher}, or polycyclic aromatic hydrocarbons \citep[PAHs;][]{1999salama,Cox11}, also in specific shapes, such as polyacenes \citep{Omontetal19}.
In the current state of research, large, carbonaceous molecules are suspected to cause most of the DIBs. This hypothesis is supported by the frequent observation of substructures in band profiles, which is generally attributed to unresolved rotational bands of heavy molecules \citep{1990A&A...233..559C, 1996A&A...307L..25E, 1993A&A...272..533E, cami2004, edibles5, edibles9}, the higher stability of large molecules against UV radiation, and the size of the only known DIB carrier C$_{60}^+$.

\begin{figure}[ht]
\centering
\subfloat[]{
\includegraphics[width=.97\hsize,trim={20 20 10 20}]{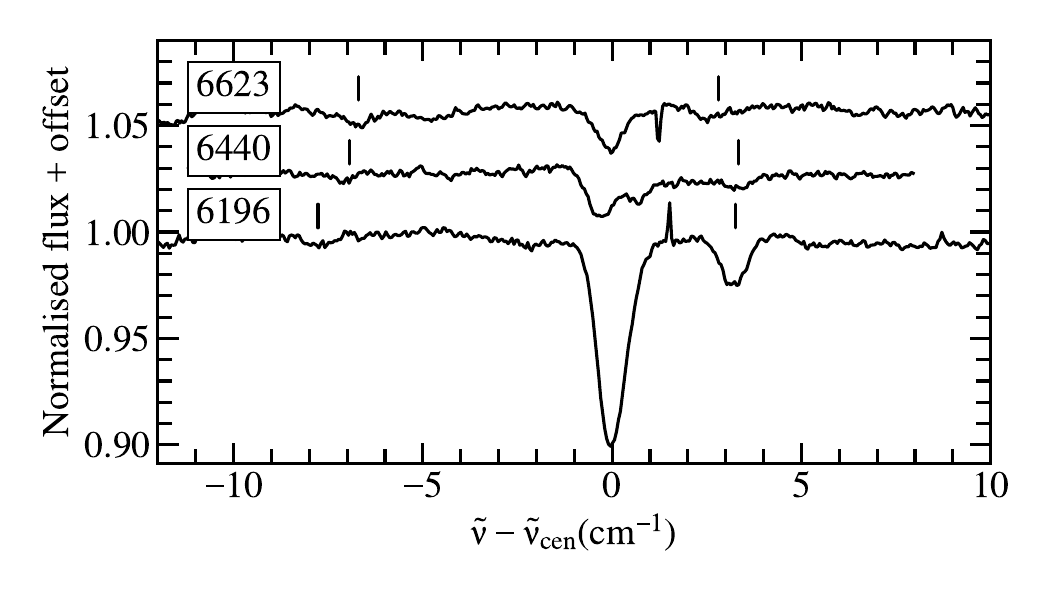}
     \label{fig:overplot_HD185418}
}\\
\subfloat[]{
\includegraphics[width=.97\hsize,trim={20 20 10 20}]{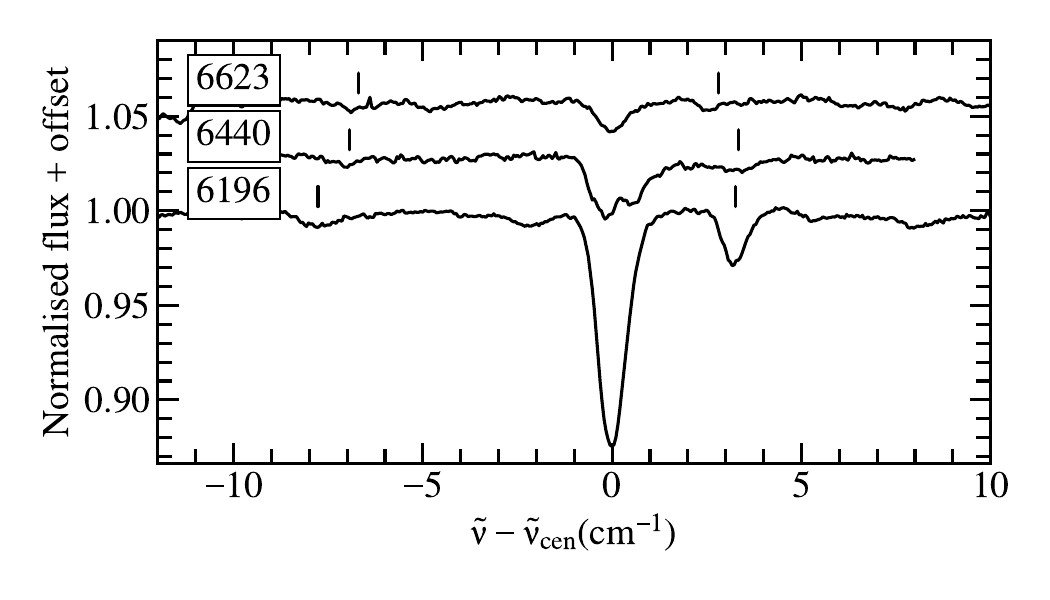}
     \label{fig:overplot_HD185859}
}\\
  \caption{DIB systems 6196, 6440, and 6623\,{\AA} in the HD~185418 (panel~\ref{fig:overplot_HD185418}) and HD~185859 (\ref{fig:overplot_HD185859}) single-cloud sight lines. The main DIBs are at $\tilde\nu = 0$ and the weaker DIBs are indicated by vertical lines. The 6440\,{\AA}~DIB is blended with another DIB. All systems were shifted to the wave number of the central DIB $\tilde\nu_\mathrm{cen}$.}
\label{fig:overplots}
\end{figure}

However, more recent work has begun to explore the possibility of smaller molecules containing transition metals as potential DIB carriers \citep{2024juanes}, potentially in cationic form.
The idea of small molecules in the interstellar medium (ISM) might be valid and worthwhile to follow up as the first molecules found in the optical spectra of the diffuse ISM such as CH \citep{Dunham37,SwRo37}, CN \citep{1940mckellar}, and CH$^+$ \citep{1941douglas} are diatomic. 
At the same time, studies of molecular ions have been carried out less frequently in the laboratory, but ionisation is ubiquitous in view of the radiation field permeating the diffuse ISM.
The search for small diatomic or maybe triatomic DIB carrier molecules has a significant advantage over the identification of larger molecules, which is due to their more thorough investigability using state-of-the-art quantum-chemical  modelling.
It allows us to compute (although still approximately) the properties  of the ground and excited electronic states of almost any pair of atoms.
This allows us to determine the properties of small DIB carriers to a very precise level if we are able to interpret interstellar spectra correctly.
As the nature of DIB carriers is still uncertain, it is important to keep in mind that the carriers probably represent a range in sizes, lifetimes, symmetries, and charges, calling for different approaches to determine the carriers responsible for each DIB class.

In this work, we explore the possibility of small, probably diatomic molecules acting as carriers of some narrow DIBs, derived purely from observational indicators. 
In Sect.~\ref{sec:observations} we introduce our spectroscopic dataset and the observational indicators, leading to the idea of small DIB carrier molecules. In Sect.~\ref{sec:methods} we describe the methods we used for calculating molecular parameters and synthetic spectra of the carriers. Based on this, we derive molecular constants and rotational temperatures in Sect.~\ref{sec:analysis}, and in Sect.~\ref{sec:discussion} we discuss the validity of the results and the differences to previous hypotheses.

\begin{table*}[t!]
\caption{DIB alignment results of the three DIB systems.} \label{tab:alignment_results}
\centering
\begin{tabular}{ccccccc}
\hline\hline
main DIB & DIB name & $\tilde\nu$ & $\Delta\tilde\nu$ & $\lambda_\mathrm{air}$ & $I$ & $r$ \\
& & (cm$^{-1}$) & (cm$^{-1}$) & ({\AA}) & &  \\
\hline
6196 & 6195 & 16138.4$\pm$0.1 & \ \ 3.2$\pm$0.1 & 6194.68$\pm$0.03 & 0.15$\pm$0.04 & 0.909 \\
 & 6196 & 16135.2$\pm$0.1 & ... & 6195.93$\pm$0.03 & 1.00$\pm$0.06 & 0.998 \\
 & 6199 & 16127.4$\pm$0.1 & $-$7.8$\pm$0.1 & 6198.92$\pm$0.05 & 0.05$\pm$0.03 & 0.649 \\
6440 & 6438 & 15528.2$\pm$0.1 & \ \ 3.4$\pm$0.2 & 6438.13$\pm$0.04 & 0.04$\pm$0.03 & 0.795 \\
 & 6440 & 15524.8$\pm$0.2 & ... & 6439.51$\pm$0.07 & 0.18$\pm$0.10 & 0.802 \\
 & 6442 & 15517.9$\pm$0.2 & $-$6.9$\pm$0.2 & 6442.39$\pm$0.06 & 0.04$\pm$0.08 & 0.173 \\
6623 & 6622 & 15098.0$\pm$0.2 & \ \ 2.8$\pm$0.2 & 6621.55$\pm$0.07 & 0.06$\pm$0.03 & 0.868 \\
 & 6623 & 15095.2$\pm$0.1 & ... & 6622.78$\pm$0.05 & 0.13$\pm$0.02 & 0.943 \\
 & 6626 & 15088.5$\pm$0.1 & $-$6.7$\pm$0.2 & 6625.72$\pm$0.05 & 0.05$\pm$0.01 & 0.922 \\
\hline
\end{tabular}
\tablefoot{Errors are calculated from standard deviations of our DIB alignment results. $\Delta\tilde\nu$ denotes the difference in wave numbers between a side DIB and its main DIB.}
\end{table*}

\section{Observations}\label{sec:observations}
\subsection{Observational data}
For this work, we used the homogeneous high-quality observational data from the ESO Diffuse Interstellar Bands Large Exploration Survey \citep[EDIBLES;][]{edibles1}. The dataset consists of spectra acquired using the Ultraviolet and Visual Echelle Spectrograph \citep[UVES;][]{Dekkeretal00} on the Very Large Telescope UT-2 at the European Southern Observatory (ESO) at Paranal, Chile. These spectra cover 123 sight lines towards Galactic O- and B-type stars. Four different instrumental  set-ups were used using the blue and red arms of UVES with a dichroic mirror to achieve a near-complete wavelength coverage between approximately 3050 and 10\,420\,{\AA}. The spectral resolving power was high, with $R$\,=\,$\lambda/\Delta\lambda$\,$\approx$\,70\,000--110\,000, utilising 0.4\arcsec \ (blue) and 0.3\arcsec \ (red) slits. The EDIBLES survey used custom procedures based on the UVES pipeline to fully reduce and calibrate the data, achieving very high signal-to-noise ratio (S/N) values of $\sim$1000 for the final data products (see \citet{edibles1} for details on target selection, observing strategy, and data reduction).

\subsection{New observational findings}
In Paper~VIII of the EDIBLES publication series \citep{ebenbichler24}, several families of DIBs with correlating band strength and profile shapes were identified from an analysis of the entire  sight line sample, called DIB profile families. 
In that paper it was concluded that DIBs of the same profile families are probably caused by the same molecules or by chemically closely related molecules.

A closer inspection of the clustering analysis from Paper~VIII in the vicinity of the three 6196\,{\AA}~DIB profile family members (the 6196, 6440, and 6623\,{\AA}~DIBs) shows that they are accompanied by weaker DIBs, only a few wave numbers apart from the central DIBs. 
Those DIBs were selected by our DIB alignment algorithm, due to their similar profile shapes compared to the 6196\,{\AA}~DIB.
This pattern was detected in a statistically significant number of sight lines, which is visualised in the histograms of Fig.~\ref{fig:hist_and_strength}, where the median strength pattern of the components is also shown, relative to the 6196\,{\AA}~DIB.\footnote{The 6196\,{\AA} DIB is the strongest among narrow DIBs, easily identifiable  even in mildly reddened sight lines, and one of the few DIBs that have been identified in other galaxies such as M31 or M33 \citep{Cordineretal08a,Cordineretal08b}.}
The information on these DIBs is summarised in Table~\ref{tab:alignment_results}, which gives for each of the three central DIBs the names of the individual component DIBs, the mean wave number $\tilde\nu$, the mean shift towards the respective central DIB $\Delta\tilde\nu$, the corresponding wavelength of the transition in air $\lambda_\mathrm{air}$, and finally the intensity ratio $I$ and the band strength Pearson correlation coefficient $r$ relative to the 6196\,{\AA}~DIB.
The radial velocities of the side and main DIBs correlate, which is indicated by the similar widths of the histograms in Fig.~\ref{fig:hist_and_strength}.
Only for the 6440\,{\AA}~DIB can a larger spread   be seen, which is expected because of the blend with another DIB.
Detection of the accompanying side DIBs strongly depends on the $S/N$ of the sight line spectra, as they can become very weak, down to a few percent of the strength of the 6196\,{\AA} DIB. We note in anticipation of later insights that a very strong correlation of these band profiles with respect to the 6196\,{\AA} DIB may not necessarily be expected because of indications for spin-orbit (\textbf{LS}) coupling\footnote{In molecular physics, the electron angular momentum pertaining to an electronic state in a molecule is denoted with \textbf{L}, and the total electronic spin angular momentum with \textbf{S}. These differ from the quantum numbers, L and S, which reflect the projection of the vectors \textbf{L} and \textbf{S} onto the molecular axis.} influencing the detailed profile shapes.

From this point on we concentrate only on a particularly valuable subset of the sight lines covered by the EDIBLES survey, the 12 single-cloud sight lines identified by \citet[their Table~1]{edibles5}.
Only for these sight lines can one single DIB carrier population with distinct environmental conditions be assumed, an assumption that is necessary for this analysis.
All central DIBs considered here and their side DIBs show the same approximately Gaussian profile shape as the 6196\,{\AA}~DIB in these single-cloud sight lines.
An example of the similarity of the three DIB systems is shown for the sight lines towards HD~185418 and HD~185859 in Fig.~\ref{fig:overplots}. We note that the 6440\,{\AA} DIB is blended with another DIB, giving rise to the double-peak form. 

In the following, we consider each of the strong DIBs with their side DIBs as a DIB system. 
Because of the systematic occurrence of the same pattern in the three DIB systems, this cannot be a coincidence. We therefore assume each DIB system to be caused by one electronic ro-vibrational transition with resolved lines.
To our knowledge, this is the first time such a correlation has been found for narrow DIBs, opening up the possibility to derive carrier properties on an observational basis, promising progress in comparison to previous attempts \citep[e.g.][]{herbig1991}.

\section{Methods}
\label{sec:methods}
We used customised Python scripts to model the electronic ro-vibrational transitions of linear molecules and fitted them to the observations using the package {\sc lmfit} \citep{newville_lmfit_2014}.
Three different kinds of electronic transitions were considered: $\Sigma\rightarrow\Sigma$, $\Sigma\rightarrow\Pi$, and $\Pi\rightarrow\Delta$.
The corresponding electronic states have the electronic angular momentum quantum number $\Lambda$\,=\,0 ($\Sigma$), $\Lambda$\,=\,1 ($\Pi$), and $\Lambda$\,=\,2 ($\Delta$).
Assuming these types of transitions, we determined the rotational constants of the electronic ground state $X$ and excited state $A$ using the positions of the three strongest transitions.
We determined the rotational constants $B''$ and $B'$ for the assumed ground and excited state, respectively, and the wave number of the electronic origin transition $\tilde\nu_0$ by evaluating Eq.~\ref{eq:rot_energy} for the respective values of the total rotational quantum numbers in the ground state ($J''$) and the excited state ($J'$) of three transitions and solving the resulting system of linear equations.
The total angular momentum must always be equal to or greater than the electronic angular momentum in the ground state ($J''$\,$\geq$\,$\Lambda''$) and the excited state ($J'$\,$\geq$\,$\Lambda'$) because it is the sum of the electronic and nuclear angular momenta.
There is never a transition directly at $\nu_0$, which would only be the case for the Q(0) transition. 
For the $\Sigma\rightarrow\Sigma$ transition no Q branch is allowed as $\Lambda' - \Lambda''$\,=\,$\Delta\Lambda$\,=\,0, and for all the other transitions either $\Lambda''$ or $\Lambda'$ is larger than 0. 
Equations \ref{eq:lin_eq_sigma_sigma}, \ref{eq:lin_eq_sigma_pi}, and \ref{eq:lin_eq_pi_delta} apply to the case of a $\Sigma\rightarrow\Sigma$, $\Sigma\rightarrow\Pi$, and $\Pi\rightarrow\Delta$ transition, respectively.
The equations may be written as
\begin{equation}
    \tilde\nu_{J\rightarrow J'} = \tilde\nu_0 + B' J'(J'+1) - B'' J''(J''+1),
\label{eq:rot_energy}
\end{equation}

% sigma -> sigma
\begin{equation}
\left( 
\begin{array}{ccc} 
1 & 2 & 0 \\ 
1 & 6 & -2 \\ 
1 & 0 & -2
\end{array} 
\right) 
\cdot 
\left(
\begin{array}{c}
\tilde\nu_0 \\
B'\\
B''
\end{array}
\right)
=
\left(
\begin{array}{c}
\tilde\nu_\mathrm{R(0)} \\
\tilde\nu_\mathrm{R(1)} \\
\tilde\nu_\mathrm{P(1)}
\end{array}
\right)
=
\left(
\begin{array}{c}
\tilde\nu_\mathrm{0\rightarrow 1} \\
\tilde\nu_\mathrm{1\rightarrow 2} \\
\tilde\nu_\mathrm{1\rightarrow 0}
\end{array}
\right),
\label{eq:lin_eq_sigma_sigma}
\end{equation}

% sigma -> pi
\begin{equation}
\left( 
\begin{array}{ccc} 
1 & 2 & 0 \\ 
1 & 6 & -2 \\ 
1 & 2 & -2
\end{array} 
\right) 
\cdot 
\left(
\begin{array}{c}
\tilde\nu_0 \\
B'\\
B''
\end{array}
\right)
=
\left(
\begin{array}{c}
\tilde\nu_\mathrm{R(0)} \\
\tilde\nu_\mathrm{R(1)} \\
\tilde\nu_\mathrm{Q(1)}
\end{array}
\right)
=
\left(
\begin{array}{c}
\tilde\nu_\mathrm{0\rightarrow 1} \\
\tilde\nu_\mathrm{1\rightarrow 2} \\
\tilde\nu_\mathrm{1\rightarrow 1}
\end{array}
\right),
\label{eq:lin_eq_sigma_pi}
\end{equation}

% pi -> delta
\begin{equation}
\left( 
\begin{array}{ccc} 
1 & 6 & -2 \\ 
1 & 12 & -6 \\ 
1 & 6 & -6
\end{array} 
\right) 
\cdot 
\left(
\begin{array}{c}
\tilde\nu_0 \\
B'\\
B''
\end{array}
\right)
=
\left(
\begin{array}{c}
\tilde\nu_\mathrm{R(1)} \\
\tilde\nu_\mathrm{R(2)} \\
\tilde\nu_\mathrm{Q(2)}
\end{array}
\right)
=
\left(
\begin{array}{c}
\tilde\nu_\mathrm{1\rightarrow 2} \\
\tilde\nu_\mathrm{2\rightarrow 3} \\
\tilde\nu_\mathrm{2\rightarrow 2}
\end{array}
\right),
\label{eq:lin_eq_pi_delta}
\end{equation}

\noindent
where $\tilde\nu_{J''\rightarrow J'}$ is the wave number of a single ro-vibrational transition from $J''$ to $J'$.
To calculate the line intensities $I(J'',J')$ of the rotational transitions from $J''$ to $J'$, we used the equation

\begin{equation}
    I(J'',J') \propto A_{\Lambda'' J''} n_{J''} = A_{\Lambda'' J''} (2J''+1) \exp{-\frac{J''(J''+1)B''}{k_B T_\mathrm{rot}}},
\label{eq:intensities}
\end{equation}

\noindent
where $k_B$ is the Boltzmann constant, $T_\mathrm{rot}$ is the rotational temperature, and $A_{\Lambda'' J''}$ is the H\"onl-London factor, as described by \citet{herzberg_molecular_1991}.
For $\Delta \Lambda$\,=\,+1, that is the $\Sigma\rightarrow\Pi$ and $\Pi\rightarrow\Delta$ transitions the H\"onl-London factors are

\begin{equation}
     A_{\Lambda'' J''} = \frac{(J'' + 2 + \Lambda'')(J'' + 1 + \Lambda'')}{(J''+1)(2J''+1)}\hspace{1cm}  \hfill(\mathrm{for}~\Delta J = +1)\hspace{0.1cm},
\end{equation}
\begin{equation}
     A_{\Lambda'' J''} = \frac{(J'' + 1 + \Lambda'')(J'' - \Lambda'')}{J''(J''+1)}\hspace{1cm}\hfill(\mathrm{for}~\Delta J = 0)\hspace{0.1cm},
\end{equation}
\begin{equation}
A_{\Lambda'' J''}  = \frac{(J'' - 1 - \Lambda'')(J'' - \Lambda'')}{J''(2J''+1)}\hspace{1cm}\hfill
     (\mathrm{for}~\Delta J = -1)\hspace{0.1cm},
\end{equation}
where $J''$ and $\Lambda''$ are the total angular momentum and the electronic angular momentum in the ground state.
In the case of the $\Sigma\rightarrow\Sigma$ transition, where $\Delta \Lambda$\,=\,0, the H\"onl-London factors are

\begin{equation}
     A_{\Lambda'' J''} = \frac{(J'' + 1 + \Lambda'')(J'' + 1 - \Lambda'')}{(J''+1)(2J''+1)}\hspace{1cm}  \hfill(\mathrm{for}~\Delta J = +1)\hspace{0.1cm},
\end{equation}
\begin{equation}
A_{\Lambda'' J''}  = \frac{(J''+ \Lambda'')(J'' - \Lambda'')}{J''(2J''+1)}\hspace{1cm}\hfill
     (\mathrm{for}~\Delta J = -1)\hspace{0.1cm}.
\end{equation}

\begin{table}[t!]
\caption{Rotational constants calculated from DIB alignment results.}
\label{tab:rot_constants}
\centering
\begin{tabular}{cccc}
\hline\hline
Transition & $B''$ & $B'$ & $\tilde\nu_0$ \\
 & (cm$^{-1}$) & (cm$^{-1}$) & (cm$^{-1}$) \\
\hline
$6196~\Sigma\rightarrow\Sigma$ & 2.05$\pm$0.07 & 1.84$\pm$0.05 & 16131.48$\pm$0.10 \\
$6196~\Pi\rightarrow\Delta$ & 1.94$\pm$0.05 & 1.84$\pm$0.05 & 16128.01$\pm$0.16 \\
$6623~\Sigma\rightarrow\Pi$ & 3.36$\pm$0.09 & 2.38$\pm$0.06 & 15090.48$\pm$0.16 \\
$6440~\Sigma\rightarrow\Pi$ & 3.47$\pm$0.12 & 2.57$\pm$0.06 & 15519.69$\pm$0.20 \\
\hline
\end{tabular}
\tablefoot{Errors are calculated with Gaussian error propagation of the measurement errors in Table \ref{tab:alignment_results}. An additional approximated systematic error of 0.4\,cm$^{-1}$ can be expected to be present, which can be caused by \textbf{LS} coupling.}
\end{table}

In the case $\Delta \Lambda$\,=\,0, $\Delta J$\,=\,0 is not allowed.
It should be noted that we assume molecules with an electron spin of $S$\,=\,0, which may be incorrect for many molecule candidates, but serves as a first approximation for the transition strengths.
For the final absorption spectrum we used a simple Gaussian as a line profile function and calculated absorption models with a linear continuum.
More detailed descriptions for the selection rules, intensity ratios of individual lines, and the $T_\mathrm{rot}$ sensitivity are presented in Appendix \ref{app:model_details}.

To find carrier candidates with the observed parameters, we performed quantum chemical calculations at the coupled clusters singles and doubles with non-iteratively included triples level with the def2QZVP basis set 
(CCSD(T)/def2QZVP optimisation of suitable molecules upon CCSD/def2TZVP pre-optimisation of all candidate molecules) to determine the rotational constants of possible diatomic $XY$ ($X$ = H, He; $Y$ = Li--Xe; and $X$, $Y$ = H--Ne) in different ionisation states with charges from $-$1 to +2 and spin multiplicity up to octuplet, as well as the HFHe$^+$ ion in the Gaussian software \citep{g16}. We note that the single-reference treatment of the coupled cluster method is potentially problematic for transition metal complexes, and should be improved upon in following investigations.
Transition metal complexes have a challenging electronic structure, and methods based on a single Slater determinant wave function might fail to describe them reliably \citep{hait_2019}.
Therefore, calculations using multi-reference methods are desirable \citep{D1CP02640B}.
Finally, the most stable molecular term was confirmed through multi-reference configuration interaction (MRCI/def2QZVP) as  implemented in Molpro \citep{Molpro1,Molpro3}. 
We note that no excitations between states of different spin multiplicity were considered; these might play a significant role in molecules including transition metals. 
We also did not consider autoionising transitions. Including these effects is outside the scope of this paper as it is a first test of this hypothesis.

\section{Analysis}
\label{sec:analysis}

\begin{figure*}[ht]
\centering
\includegraphics[width=.98\hsize]{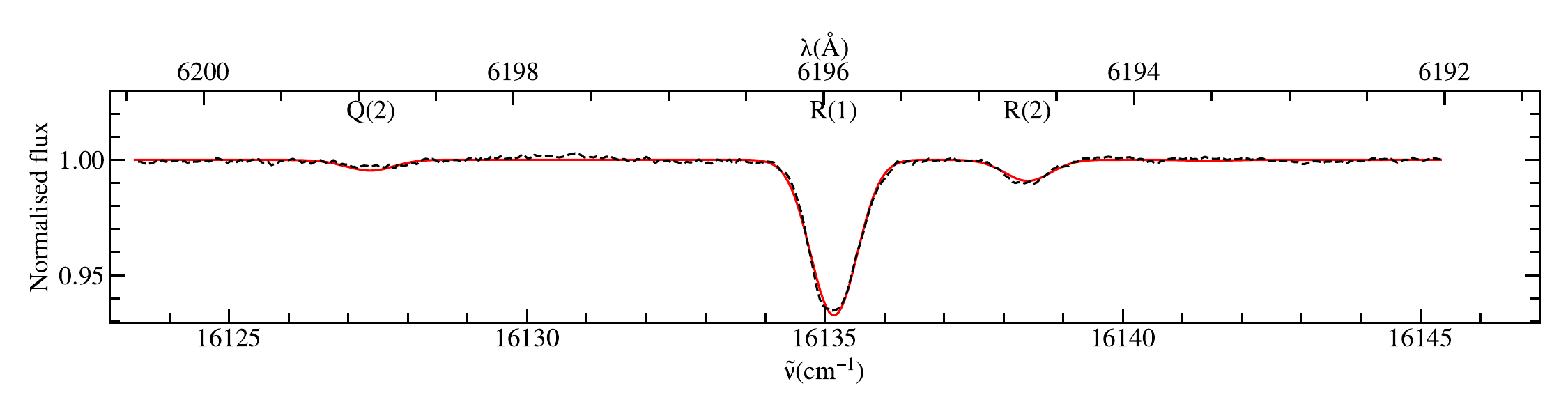}
  \caption{Fit of the 6196\,{\AA} DIB system in the HD~170740 sight line (dashed black)  using our $\Pi\rightarrow\Delta$ transition model (red).}
     \label{fig:HD170740_6196}
\end{figure*}

% Calculating rotational constants
In the 6196\,{\AA}~DIB system, the  lower-energy DIB is weaker than the central DIB, and the higher-energy DIB is even weaker (Fig.~\ref{fig:hist_and_strength_6196}). 
In order to generate a ro-vibrational transition of this pattern, either an electronic $\Sigma\rightarrow\Sigma$ (Fig.~\ref{fig:HD185859_fit}) or a $\Pi\rightarrow\Delta$ (Fig.~\ref{fig:HD170740_6196}) transition is required.
For the 6440 and 6623\,{\AA}~DIB systems, the side peaks are of equal strengths within the errors (i.e. $I(6438)$\,=\,$I(6442)$, $I(6622)$\,=\,$I(6626)$), but weaker than the central DIB (see Figs.~\ref{fig:hist_and_strength_6440} and \ref{fig:hist_and_strength_6623}, or Table~\ref{tab:alignment_results}). 
This pattern can only be reproduced by an electronic $\Sigma\rightarrow\Pi$ transition, where $I($R$(1))$\,=\,$I($Q$(1))$.
We determined the molecular constants for each transition scenario by solving the linear Eqs.~\ref{eq:lin_eq_sigma_sigma}, \ref{eq:lin_eq_sigma_pi}, and \ref{eq:lin_eq_pi_delta} for the transitions listed in Table~\ref{tab:alignment_results}. 
The results are shown in Table~\ref{tab:rot_constants}, which lists $B''$, $B'$, and $\tilde\nu_0$ for the four transition~scenarios~investigated.
The systematic errors are derived from the standard deviations of the DIB alignment results, and a systematic error is caused by possible \textbf{LS} coupling, as described below.

\begin{figure*}[ht]
\centering
\includegraphics[width=.98\hsize]{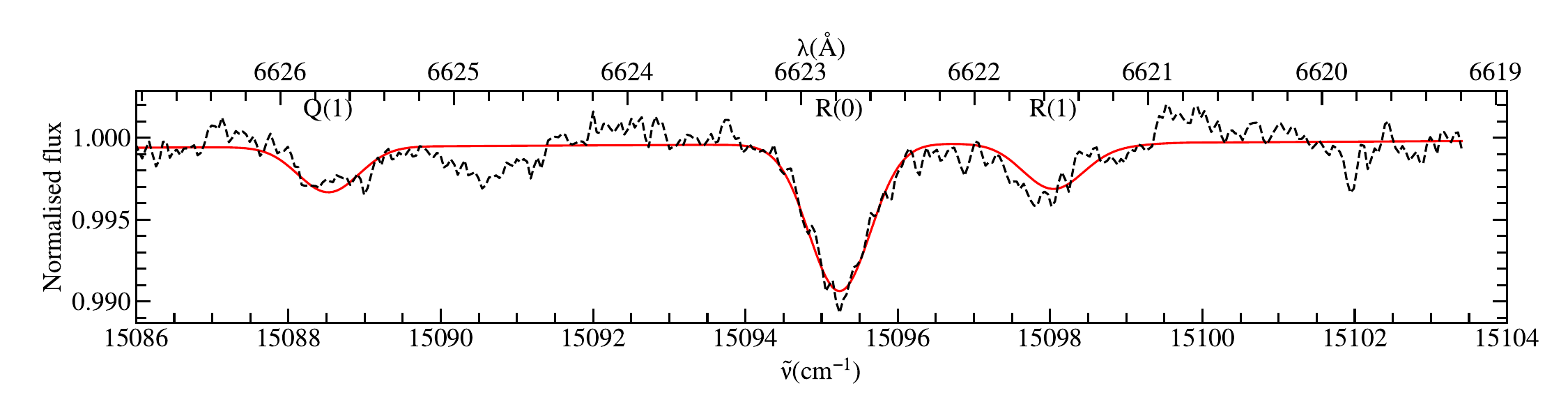}
  \caption{Fit of the 6623\,{\AA}~DIB system in the HD~170740 sight line (dashed black) using our $\Sigma\rightarrow\Pi$ transition model (red).}
     \label{fig:HD170740_6623}
\end{figure*}

% Fitting rotational temperatures
As a test of our ro-vibrational transition model, we fitted the 6196 and 6623\,{\AA}~DIBs and their respective side bands using the molecular model only by varying the rotational temperature $T_\mathrm{rot}$, the width of a Gaussian profile $\sigma$, and the radial velocity $RV$ for our single-cloud sight lines. 
The 6440\,{\AA}~DIB could not be fully investigated because of its strong blend with another DIB.
However,  considering the determined line ratios, that is two equally strong secondary DIBs, the most probable transition type is a $\Sigma\rightarrow\Pi$ transition, just as for the 6623\,{\AA}~DIB system.
The results of the 6196\,{\AA}~DIB system are shown in Table \ref{tab:fit_results_6196} and of the 6623\,{\AA}~DIB system in Table \ref{tab:fit_results_6623}.
The (discarded) results of the 6196\,{\AA}~DIB system for a $\Sigma\rightarrow\Sigma$ transition are shown in Table~\ref{tab:fit_results_6196_sigma_sigma} together with a plot of the fit in the HD~170740 sight line in Fig.~\ref{fig:HD185859_fit}.
We show our adopted fit for the HD~170740 sight line for the 6196\,{\AA}~DIB system in Fig.~\ref{fig:HD170740_6196} and for the 6623\,{\AA}~DIB system in Fig.~\ref{fig:HD170740_6623}.
The other fits for our single-cloud sight lines are shown in Figs.~\ref{fig:6196_fits_1}, \ref{fig:6196_fits_2}, and \ref{fig:6623_fits}. We note that the other five sight lines have S/N values that are too low to be analysed for the 6623\,{\AA} DIB system.
The use of a Gaussian profile is sufficient for our purpose in describing the individual components.
The next weaker rotational components R(3) and Q(3) are too weak to be detectable in any sight line.

\begin{table}[t!]
\caption{Fit results of the 6196\,{\AA}~DIB system using a $\Pi\rightarrow\Delta$ transition model.} \label{tab:fit_results_6196}
\centering
\begin{tabular}{lccr}
\hline\hline
Sight line & $T_\mathrm{rot}$ & $\sigma$ & $RV$ \\
 & (K) & (km\,s$^{-1}$) & (km\,s$^{-1}$) \\
\hline
\object{HD23180} & 4.96$\pm$0.15 & 7.71$\pm$0.10 & 12.56$\pm$0.09 \\
\object{HD24398} & 5.86$\pm$0.13 & 7.45$\pm$0.08 & 13.20$\pm$0.08 \\
\object{HD144470} & 4.55$\pm$0.21 & 8.79$\pm$0.14 & $-$9.66$\pm$0.14 \\
\object{HD147165} & 4.05$\pm$0.34 & 9.86$\pm$0.24 & $-$7.67$\pm$0.24 \\
\object{HD147683} & 5.27$\pm$0.21 & 8.02$\pm$0.14 & $-$0.99$\pm$0.14 \\
\object{HD149757} & 5.27$\pm$0.23 & 9.63$\pm$0.18 & $-$14.53$\pm$0.18 \\
\object{HD166937} & 4.46$\pm$0.13 & 7.68$\pm$0.08 & $-$6.34$\pm$0.08 \\
\object{HD170740} & 5.24$\pm$0.07 & 7.16$\pm$0.04 & $-$11.37$\pm$0.04 \\
\object{HD184915} & 5.38$\pm$0.15 & 6.88$\pm$0.09 & $-$12.62$\pm$0.08 \\
\object{HD185418} & 5.83$\pm$0.12 & 6.95$\pm$0.07 & $-$9.02$\pm$0.07 \\
\object{HD185859} & 5.62$\pm$0.09 & 6.48$\pm$0.05 & $-$8.50$\pm$0.05 \\
\object{HD203532} & 4.52$\pm$0.16 & 7.27$\pm$0.09 & 13.94$\pm$0.09 \\

\hline
\end{tabular}
\tablefoot{The radial velocity ($RV$) is measured in the barycentric rest frame.}
\end{table}

\begin{table}[t!]
\caption{Fit results of the 6623\,{\AA}~DIB system using a $\Sigma\rightarrow\Pi$ transition model.} \label{tab:fit_results_6623}
\centering
\begin{tabular}{lccr}
\hline\hline
Sight line & $T_\mathrm{rot}$ & $\sigma$ & $RV$ \\
 & (K) & (km\,s$^{-1}$) & (km\,s$^{-1}$) \\
\hline
HD23180 & 5.47$\pm$1.03 & 6.03$\pm$0.91 & 14.40$\pm$0.87 \\
HD24398 & 6.43$\pm$0.62 & 10.12$\pm$0.86 & 13.32$\pm$0.79 \\
HD170740 & 6.11$\pm$0.25 & 7.94$\pm$0.28 & $-$11.01$\pm$0.27 \\
HD184915 & 6.09$\pm$0.58 & 6.01$\pm$0.48 & $-$11.46$\pm$0.46 \\
HD185418 & 5.50$\pm$0.29 & 6.70$\pm$0.29 & $-$7.95$\pm$0.28 \\
HD185859 & 4.76$\pm$0.25 & 7.20$\pm$0.28 & $-$8.35$\pm$0.27 \\
HD203532 & 6.84$\pm$0.74 & 6.94$\pm$0.65 & 13.61$\pm$0.62 \\
\hline
\end{tabular}
\tablefoot{If a sight line is not listed, the fit was not successful due to the low signal-to-noise ratio. $RV$ is measured in the barycentric rest frame.}
\end{table}

\begin{figure*}[ht]
\centering
\includegraphics[width=.98\hsize]{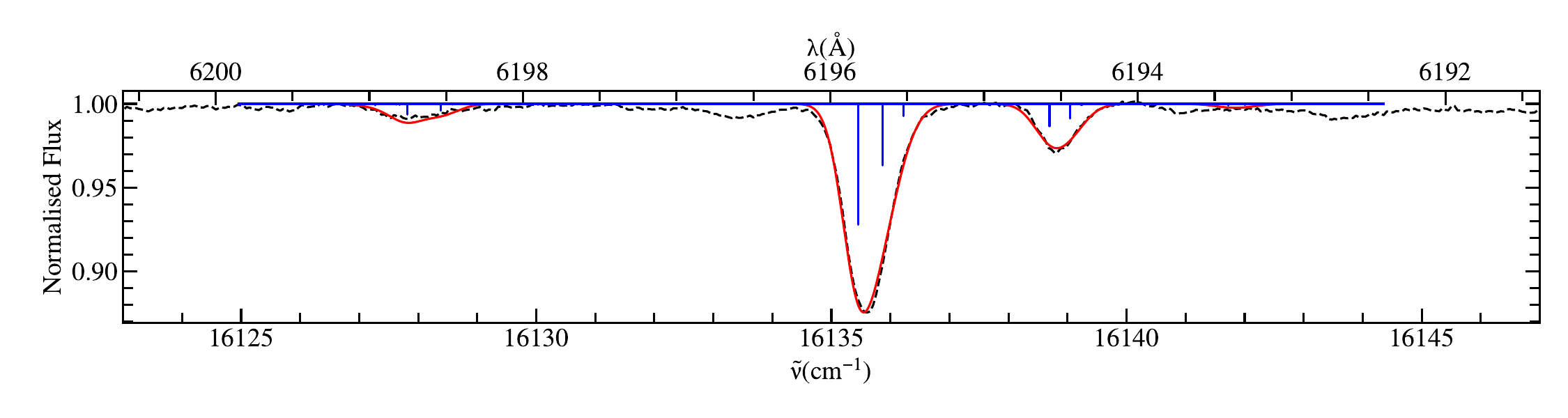}
  \caption{ 6196\,{\AA} DIB system in the HD~185859 sight line (dashed black)  and a $^2\Pi\rightarrow\, ^2\Delta$ transition model including \textbf{LS} coupling (red), generated with Pgopher. The unconvolved `stick' spectrum is shown in blue. This plot shows how \textbf{LS} coupling can explain the systematic asymmetry of the 6196\,{\AA}~DIB.}
     \label{fig:ls_fit}
\end{figure*}

For the 6196\,{\AA}~DIB system, our data had the necessary S/N to perform an additional analysis of the DIB profile shapes.
On closer inspection of the profiles, one finds that the components of the 6196\,{\AA}~DIB system have asymmetric profile shapes, which is systemic in the single-cloud sight lines.
For a $^2\Pi\rightarrow\, ^2\Delta$ transition, \textbf{LS} coupling (which is expected for such a transition) can explain this asymmetry.
We generated a model with Pgopher \citep{pgopher}, using the parameters $B''$\,=\,1.9\,cm$^{-1}$, $B'$\,=\,1.8\,cm$^{-1}$, and \textbf{LS} coupling constants in the ground ($A''$) and excited state ($A'$) of $A''$\,=\,$A'$\,=\,$-$0.45\,cm$^{-1}$.
Due to the shifts in \textbf{LS} coupling, we had to change the rotational constants by 0.4\,cm$^{-1}$ for each electronic level.
This is an additional systematic error source of the values in Table \ref{tab:rot_constants}.
As a first approximation, we added a systematic error of 0.4\,cm$^{-1}$ to the statistical error from DIB alignment.
The comparison of the model with the 6196\,{\AA}~DIB system in the HD~185859 sight line is displayed in Fig.~\ref{fig:ls_fit}.
The effect of the \textbf{LS} coupling has almost the identical effect on the 6195\,{\AA}~DIB and causes a stronger split for the 6199\,{\AA}~DIB.
This \textbf{LS} coupling improves the fit overall and can explain the systematically asymmetric profile shape of the DIB system.
We abstained from implementing and fitting a detailed molecular model with \textbf{LS} coupling and leave this qualitative explanation for the profile shapes, as the overall picture will be little changed.

With this hypothesis, supported by the observed molecular constants and the types of electronic transitions, we searched for appropriate carrier candidates for our three DIB systems.
We compared the rotational constants of stable molecules calculated in Gaussian with our observations in Table \ref{tab:rot_constants}.
If the theoretical value matched the observation within 20\% and the relative energy of the respective electronic state was within 0.6\,eV with respect to the tentative most stable spin multiplicity, we checked if the molecule candidates had excited states above the ground state with an energy difference close to the origin wave number $\tilde\nu_0$, that is in the 1.5--3\,eV region, as well as the required symmetry of the molecular terms. We also excluded some molecules and ions for which experimental measurements already provided rotational constants or excitation energies incompatible with those given in Table \ref{tab:rot_constants}.
The resulting candidates are listed in Table \ref{tab:carrier_candidates}.

%--------------------------------------------------------------------
\section{Discussion}
\label{sec:discussion}
\subsection{Previous interpretations of the 6196\,{\AA}~DIB}
The assumption of a small polar molecule as the carrier of the 6196\,{\AA}~DIB is a strong contradiction to previous work, which assumed a large carrier molecule, due to the presence of substructures. This relies on the finding of two subpeaks separated by $\sim$8\,km\,s$^{-1}$ in the 6196\,{\AA}~DIB towards the \object{Sco OB2} association \citep{galazutdinov2023}.
This substructure was interpreted as potential PR-branches of a very large molecule with a very small rotational constant $B$.
However, a study of neutral hydrogen by \citet{1974Isancisi} finds an expanding \ion{H}{i} shell associated with Sco OB2, 
which could be the result of an old supernova remnant, slowed down by interaction with the ISM.
Its expansion speed is 3\,km\,s$^{-1}$ in each direction.
The observed substructures could therefore simply be different velocity components of the same carrier concentrated on the shell rim, traversed twice by the line of sight.
\citet{2009A&A...498..785K} report a correlation between the rotational temperature of interstellar C$_2$ and the full width at half maximum of the 6196\,{\AA}~DIB, indicating that the DIB is caused by a large molecule and shows unresolved rotational bands.
However,  the sample size is too small to be certain about the correlation, and more importantly the 6196\,{\AA}~DIB is a $\sigma$ DIB, so its carrier resides in very different cloud environments than C$_2$.
Hence, this correlation must be taken with caution.

However, the substructure could not be systematically confirmed in the EDIBLES single-cloud sight lines at $R$\,$\approx$\,110\,000, unlike in the case for the 5797, 6397, or 6614\,{\AA}~DIBs \citep{edibles5}.
We fitted a double Gaussian to the 6196\,{\AA}~DIB and inspected the residuals to see small-scale structures. 
We did not find any intrinsic patterns that exceed the noise level, which is illustrated in Fig.~\ref{fig:double_gauss_fits}.
In addition,  no substructure of the 6196\,{\AA}~DIB was  found initially at an instrumental resolution of 0.075\,{\AA}, which equates to $R$\,=\,$\lambda / \Delta \lambda$\,$\approx$\,80\,000 \citep{smith1981}, which was confirmed by \citet{herbig1982}.

In other work, DIB studies on the \object{Herschel 36} sight line found no substructure \citep{oka2013}, but extended red wings for multiple DIBs, also for the 6196\,{\AA}~DIB. 
This finding is in conflict with the hypothesis of a large carrier molecule showing rotational splitting because the temperature in the Herschel~36 sight line is assumed to be rather high and  does not show splitting.
The extended red wings can be explained by the Doppler shift as well, since strong dynamics and large  redshifts of $\sim$80\,km\,s$^{-1}$ of H$_2$ in the vicinity of Herschel 36 were observed by \citet{2014rachford}.
The red wings may not be visible in other molecules and ionised species because they reside in different areas of the ISM.
These dynamics could also influence the DIB structure in this particular sight line.
In summary, it has not been unambiguously proven that the 6196\,{\AA}~DIB shows systematic substructure, so the possibility of a smaller molecular carrier should not be ignored.

\begin{table}[t]
    \caption{Rotational constants of proposed carrier molecules as calculated at the CCSD(T)/def2QZVP level, along with the most stable molecular term as determined at the MRCI/def2QZVP level.}
    \centering
    \begin{tabular}{lccl}
    \hline\hline
    Molecule & $B''$ & Molecular & Lit.\\
    & (cm$^{-1}$) & ground term &  \\
    \hline
    (NO$^{2+}$) & 1.93 & $^2\Sigma^+$ & o\tablefootmark{a}\\
    (HFHe$^+$) & 1.94 & $^2\Pi$ & ...\\
    \hline
    KH$^{-}$ & 2.99 & $^2\Sigma^+$ & ...\\
    SrH$^{-}$ & 3.31 & $^1\Sigma^+$ & ...\\
    SrH & 3.66 & $^2\Sigma^+$ & f\tablefootmark{b}\\
    (CrH$^{+}$) & 3.74 & $^7\Sigma^+$ & o\tablefootmark{c}\\
    CaH$^{-}$ & 3.85 & $^1\Sigma^+$ & ...\\
    \hline
    \end{tabular}
    \label{tab:carrier_candidates}
    \tablebib{
    \tablefoottext{a} {\citet{2008JChPh.128n4301B}}; 
    \tablefoottext{b} {\citet{2016JQSRT.170..169L}}; 
    \tablefoottext{c} {\citet{2022JPCM...34K4001R}} 
    }
    \tablefoot{Molecules in the top part are candidates for the 6196\,{\AA}~DIB system; molecules in the bottom part are candidates for the other two systems. The species in brackets are metastable. In the last column (Lit.), molecules marked with `f' are ruled out as carriers, `...' marks the absence of literature, and `o' marks molecules for which there is literature available, but the carrier is neither confirmed nor ruled out.}
\end{table}

\subsection{A new paradigm for the 6196\,{\AA}~DIB}
In a change of approach, we assume that the profile of the 6196\,{\AA}~DIB and of its profile family members are purely caused by Doppler shifts.
The fitted Doppler width of $\sigma$\,$\approx$\,7\,km\,s$^{-1}$ is considerably larger than the Doppler width of most known interstellar atomic and molecular lines. 
An explanation for this phenomenon might be that those DIB carriers might not be stable against elastic collisions, which are needed to reduce the turbulent velocity dispersion in interstellar gases.
The carriers would only exist with velocity dispersion originating from their production,  a velocity dispersion that can never narrow down.
Hence, this would imply a continuous formation and destruction of DIB carriers.
The perceived velocity distribution could be caused during the production of the carrier molecule, which may be influenced by the high levels of turbulence in the diffuse ISM.
The 6196\,{\AA}~DIB is strongly correlated with the 6614\,{\AA}~DIB \citep{McCall2010ApJ...708.1628M}, for which \cite{edibles9} reported correlating line widths with CH$^+$.
According to \citet{2023A&A...669A..74G}, CH$^+$ is the consequence of mass exchanges between the cold neutral medium (CNM) and the warm neutral medium (WNM), induced by large-scale turbulence and thermal instability, indicating that large line widths are likely caused by turbulent environments.
We expect a variable line width of the 6196\,{\AA}~DIB across different sight lines due to this effect.
The timescale of the formation destruction cycle of the carrier could be comparable to CH$^+$ \citep[their Table 2]{2023A&A...669A..74G}, but at this point we are not able to provide reliable estimates.
A molecular broadening mechanism  such as lifetime broadening is very unlikely, because of the sharpness of some subfeatures in the `split' sight lines towards the Sco~OB2 association.
Moreover, we focus on each DIB system as a whole, rather than specifically trying to explain the detailed profile shape of the 6196\,{\AA}~DIB itself.

The determined molecular constants point to small linear molecules as the carriers of the investigated DIB systems.
Because of the large rotational constant, most likely they are diatomic, possibly triatomic.
Due to the different $B''$, the 6196~{\AA}~DIB system must have a different carrier than the other two, while the 6440 and 6623~{\AA}~DIB systems could have the same carrier.
The fitted rotational temperatures (Tables \ref{tab:fit_results_6196}, \ref{tab:fit_results_6623} and \ref{tab:fit_results_6196_sigma_sigma}) show consistently low temperatures of less than 7\,K, making polar molecules much more likely than nonpolar carriers.
For the 6196\,{\AA}~DIB system, the $\Sigma\rightarrow\Sigma$ transition model shows rotational temperatures below the CMB temperature for all of our single-cloud sight lines.
Such low temperatures are very unlikely. 
A polar molecule should have at least the CMB temperature of $T_\mathrm{CMB}=2.7260\pm0.0013$\,K \citep{Fixsen_2009} because the CMB is the strongest radiation source in the diffuse ISM at radio wavelengths \citep[see e.g.][Chapter~12]{Draine11}.
The $\Pi\rightarrow\Delta$ model shows rotational temperatures $T_\mathrm{rot}$\,$>$\,$T_\mathrm{CMB}$ for all sight lines, as well as the fits for the 6623\,{\AA}~DIB system.
The $T_\mathrm{rot}$ values of the 6196 and 6623\,{\AA}~DIB systems do not correlate with each other, but both range between 4 and 7\,K.
These temperatures are close to the known rotational temperatures of interstellar polar diatomics.
Measurements of CO in the UV range by \citet{1982wannier} show a rotational temperature of CO of $T_\mathrm{rot}$\,=\,4\,K in the HD~149757 ($\zeta$~Oph) sight line.
It is a promising outcome of our analysis that we determine rotational temperatures in this range for a candidate polar diatomic carrier.

The search for such a carrier proved to be difficult.
Only a handful of diatomics showed rotational constants in the right ranges and additionally have the right electronic states to match our observations (see Table \ref{tab:carrier_candidates}).
The list contains a doubly ionised molecule, which seems very exotic at first sight, but this scenario should not be discarded in the harsh ISM environment.
Although the doubly ionised species are only metastable against dissociation, we still list them as candidates because they can still cause DIBs if their abundance is high enough.
It might be possible that some DIB carriers are short-lived, but are produced at a sufficient rate because many molecules cannot survive the harsh interstellar radiation field.
Hence, collision complexes could be possible carriers of some DIBs. 

Furthermore, we also consider the abundances of the constituents of the molecules to assess whether they may exist in sufficient quantities to produce the observed DIBs. We
use the  present-day cosmic abundances \citep{NiPr12,Przybillaetal08b,Przybillaetal13}, supplemented by solar abundances \citep[e.g.][]{2021A&A...653A.141A} for the missing species.

% NO2+
A candidate in Table~\ref{tab:carrier_candidates} is NO$^{2+}$, but it has a $\Sigma$ ground state, and the excited electronic state closest to 16130\,cm$^{-1}$ is a $\Pi$ state \citep{2008JChPh.128n4301B}, which is incompatible with our observations of the 6196\,{\AA}~DIB system.
However, the A$^2\Pi$ state is expected to be heavily blended with the X$^2\Sigma$ state from simple considerations of energetic proximity and perturbation selection rules.
The biggest caveat with this methodology is that it does not account for the nuances in the excited states themselves, namely that of state mixings or perturbations.
Technically, this means that we cannot completely discount NO$^{2+}$ on the basis of this argument, especially since nitrogen and oxygen are both amongst the most abundant elements with $\epsilon(\mathrm{N})=\log(\mathrm{N/H})+12=7.79\pm0.04$ and $\epsilon(\mathrm{O})=8.76\pm0.05$.

We found no studies regarding possible interstellar abundances for KH$^-$, SrH$^-$, or CaH$^-$, and we can only discuss the abundances of the heavy molecular components.
Potassium and calcium are quite abundant ($\epsilon(\mathrm{K})=5.08\pm0.04$, $\epsilon(\mathrm{Ca})=6.29\pm0.03$), which means that some amount of both molecular anions may be present in the ISM.
Strontium is rather rare ($\epsilon(\mathrm{Sr})=2.88\pm0.03$), but it is among the first-peak elements of the s-process. Consequently, the circumstellar envelopes' asymptotic giant branch (AGB) may be the formation sites of the neutral molecules favoured by the enhanced s-process abundances and generally higher densities there, later to be further processed in the ISM. 

% SrH
SrH is one of the most intensely studied molecules in our candidate list. Both experimental and theoretical measurements \citep{2016JQSRT.170..169L} disagree with our observed ground-state rotational constant, and thus confirm that SrH can be excluded as a candidate due to a mismatch of rotational constants and electronic energy levels.
% CrH+
Although it is probably metastable with a relative energy of about 1\,eV, one interesting candidate is CrH$^+$, because molecular CrH has been observed in sunspots \citep{1980A&AS...42..209E}, S-type stars \citep{1980A&A....84..300L}, and brown dwarfs \citep{1999ApJ...519..834K}.
In a theoretical study, \citet{2022JPCM...34K4001R} found that CrH$^+$ can be formed by electron ionisation of CrH, whereas the cation has a smaller ionisation cross-section than the neutral.
It was also found that CrH$^+$ is dissociated upon recombination with an electron, making it more stable in ISM regions of very low density, such as the DIB environment.
The abundance of Cr is also rather high ($\epsilon(\mathrm{Cr})=5.63\pm0.02$), making CrH$^+$ more likely as a carrier.
Generally, metal-hydride cations could be good candidates as similar molecules have been detected in the ISM, such as ArH$^+$ \citep{2013Sci...342.1343B}.
For completeness, we examined deuteride molecules and ions as candidates, which provide a larger number of candidate carriers than hydride ions, but which can be expected to be significantly rarer because of the low deuterium-to-hydrogen ratio.
The discussion can be found in Appendix~\ref{app:deuterated}. 
The anions HK$^-$, SrH$^-$, and CaH$^-$ are not intuitive because of their negative charge, but they should not be excluded because of this argument.

As one of the tri-atomic carrier candidates, HFHe$^+$ has a rotational constant that is very close to our observed value for the 6196\,{\AA} DIB.
This complex is  metastable, and would have to be prepared in an excited state before fluorescing to the ground state, but HF has been found to be abundant at the edge of interstellar clouds, for example in the Orion Bar \citep{2019A&A...631A.117K}.
HF should be the dominant reservoir of interstellar fluorine, due to the exothermic reaction $\mathrm{H}_2 + \mathrm{F} \rightarrow \mathrm{HF} + \mathrm{H}$.
Consequently, $\mathrm{HF} + \mathrm{He}^+ \rightarrow \mathrm{HFHe}^+$ could be formed.
In theoretical studies, the interaction of (HF$^+$) and He has been studied on multiple occasions because of the significance of HF in laser physics.
The newest study by \citet{2024JQSRT.32809168N} concludes that the configuration FHHe$^+$ is more strongly bound than HFHe$^+$ and also has a much smaller $B''$.
This is consistent with previous studies, as shown by \citet{1994CPL...220..117S} and \citet{2004JChPh.120...93L}.
However, HF is abundant at the edge of the Orion Bar, coinciding with the skin-effect\footnote{Many DIBs -- especially $\sigma$ type -- are primarily found along the edge or skin of interstellar clouds. This is known as the skin-effect.} phenomenon of DIBs.
This makes HFHe$^+$ a good candidate from the observational point of view.

Using white dwarfs and Extreme Ultraviolet Explorer (EUVE) data, \citet{1999A&A...346..969W} found that He is significantly ionised within the first one hundred parsecs around the Sun, with $\mathrm{He}^+ / (\mathrm{He} + \mathrm{He}^+)$\,$\approx$\,40\%.
This high degree of ionisation was recently confirmed, at least for the nearby clouds, on the basis of the interstellar flow within the  Solar System \citep{2019ApJ...882...60B}.
Recently, \citet{2019Natur.568..357G} have detected HeH$^+$ in a planetary nebula.
Some evidence of helium-containing DIB carriers might even have been observed in the laboratory in plain sight by \citet{linnartz2010}.
Unfortunately, the carrier could not be unambiguously identified in that work, due to the short lifetime of the transition of $\tau$\,$\approx$\,0.15\,ps, but it certainly contains hydrogen and carbon. 
As a standard procedure, but interestingly in this context, a $1\%$ C$_2$H$_2$/He mixture was used for the experiment. 
Normally, He is assumed to not react with other elements or molecules, but in a plasma it could be ionised, thus opening up several reaction channels. 
This would explain the absence of the feature without a plasma. 
Given the high abundance of He in the ISM, a significant fraction should be ionised to He$^+$, especially in the presumably extreme DIB environment, potentially forming cationic molecules with other abundant atoms.

\section{Conclusions}
\label{sec:conclusions}
In this work we have shown that there is a consistent pattern of the 6196, 6440, and 6623\,{\AA}~DIBs and their respective weaker side DIBs.
The hypothesis of rotational bands being the cause is a first approach to explaining these patterns and supplies a list of specific small carrier molecule candidates that are testable with more detailed theoretical calculations, with additional observations, and finally in the laboratory.
Other mechanisms that are able to cause this pattern, such as spin-orbit coupling, should also be addressed  in future studies.
In the coming years, the computational precision in quantum chemistry will only increase, giving us the possibility to search for small carrier candidates more precisely.
In broader theoretical studies of diatomic and maybe triatomic molecules, it should be possible to determine a DIB carrier with the information shown here.
To investigate this hypothesis further, detailed profile fitting of whole DIB systems has to be implemented, which will constrain additional molecular constants and clarify whether the hypothesis is valid.
If such small DIB carriers exist, they should have rotational transitions in the electronic ground state as well.
Comprehensive searches for such transitions in the radio, in emission, or in absorption are necessary.
In this work we have shown that small molecules could very well be good DIB carrier candidates, with many observational indicators supporting this assumption.
We want to emphasise the idea of collision complexes as possible DIB carrier candidates such as HFHe$^+$.
Because of the extreme conditions in the diffuse ISM, chemical processing could be enhanced, possibly leading to abundances of such complexes that are high enough to become observable.
It is important for the progress in DIB research to keep an open mind concerning all possible theories.

\begin{acknowledgements}
A.E.~and N.P.~gratefully acknowledge support from the Austrian Science Fund FWF, Grant DOI 10.55776/W1259                                (DK-ALM). JC acknowledges support from an NSERC Discovery Grant.
This article/publication is based upon work from COST Action CA21126 - Carbon molecular nanostructures in space (NanoSpace), supported by COST (European Cooperation in Science and Technology).
This work is based on observations collected at the European Organisation for Astronomical Research in the Southern Hemisphere under ESO programme 194.C-0833.  
Data are publicly available through the ESO Science Archive Facility at \url{http://archive.eso.org/eso/eso_archive_main.html}. The computational results presented were achieved using the HPC infrastructure LEO of the University of Innsbruck.
\end{acknowledgements}

% WARNING
%-------------------------------------------------------------------
% Please note that we have included the references to the file aa.dem in
% order to compile it, but we ask you to:
%
% - use BibTeX with the regular commands:
%   \bibliographystyle{aa} % style aa.bst
%   \bibliography{Yourfile} % your references Yourfile.bib
%
% - join the .bib files when you upload your source files
%-------------------------------------------------------------------

\bibliography{bibliography} % your references Yourfile.bib
\bibliographystyle{aa} % style aa.bst
\listofobjects

% \onecolumn
\begin{appendix}
\section{Discarded \texorpdfstring{$\Sigma\rightarrow\Sigma$}\ ~transition model for the 6196\,{\AA} DIB system}
\label{app:sigma_sigma}
\begin{figure}[ht]
\centering
\includegraphics[width=\hsize]{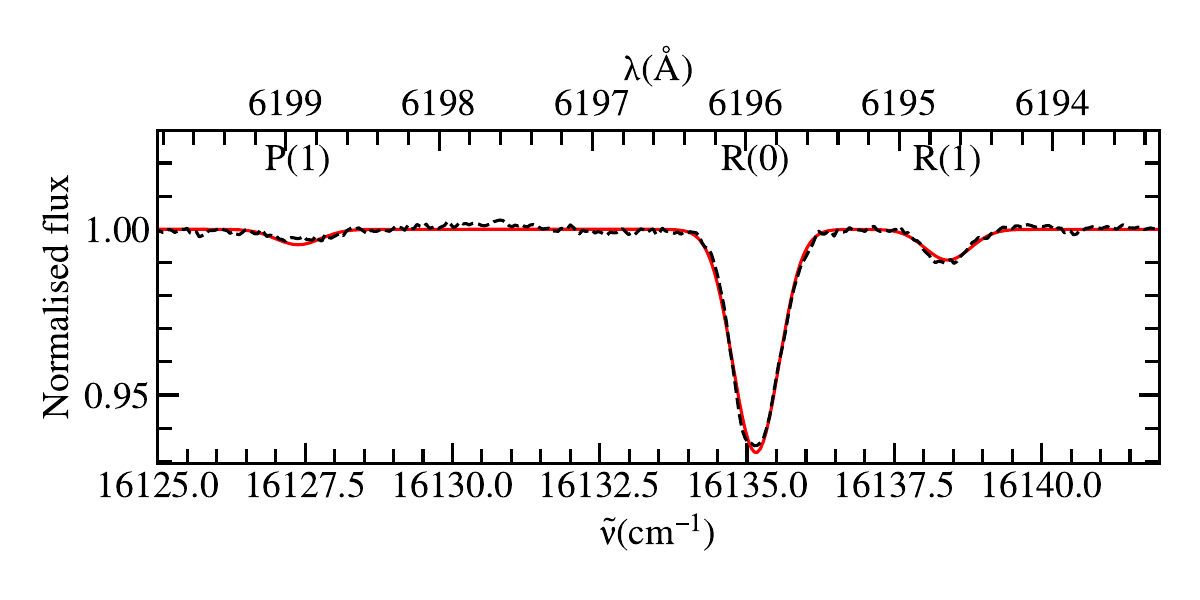}
  \caption{6196\,{\AA}~DIB and its companions in the single cloud sight line towards HD~170740 (dashed black) compared with a synthetic spectrum (in red) of a $\Sigma\rightarrow\Sigma$ transition of a linear molecule with $T_\mathrm{rot}=2.18\pm 0.02$\,K.
          }
     \label{fig:HD185859_fit}
\end{figure}

\begin{table}[th]
\caption{Fit results for the single cloud dominated 6196\,{\AA}~DIB sightlines using a $\Sigma\rightarrow\Sigma$ transition model.} \label{tab:fit_results_6196_sigma_sigma}
\centering
\begin{tabular}{lccr}
\hline\hline
Sight line & $T_\mathrm{rot}$ & $\sigma$ & $RV$ \\
 & (K) & (km\,s$^{-1}$) & (km\,s$^{-1}$) \\
\hline
HD23180 & 2.08$\pm$0.05 & 7.707$\pm$0.097 & 12.56$\pm$0.09 \\
HD24398 & 2.41$\pm$0.04 & 7.456$\pm$0.079 & 13.20$\pm$0.08 \\
HD144470 & 1.94$\pm$0.07 & 8.791$\pm$0.144 & $-$9.67$\pm$0.14 \\
HD147165 & 1.76$\pm$0.12 & 9.854$\pm$0.244 & $-$7.67$\pm$0.24 \\
HD147683 & 2.18$\pm$0.07 & 8.009$\pm$0.140 & $-$0.99$\pm$0.14 \\
HD149757 & 2.20$\pm$0.08 & 9.629$\pm$0.181 & $-$14.53$\pm$0.17 \\
HD166937 & 1.90$\pm$0.05 & 7.675$\pm$0.082 & $-$6.34$\pm$0.08 \\
HD170740 & 2.18$\pm$0.02 & 7.162$\pm$0.043 & $-$11.36$\pm$0.04 \\
HD184915 & 2.22$\pm$0.05 & 6.879$\pm$0.085 & $-$12.62$\pm$0.08 \\
HD185418 & 2.37$\pm$0.04 & 6.943$\pm$0.067 & $-$9.02$\pm$0.07 \\
HD185859 & 2.31$\pm$0.03 & 6.473$\pm$0.049 & $-$8.50$\pm$0.05 \\
HD203532 & 1.93$\pm$0.06 & 7.273$\pm$0.094 & 13.94$\pm$0.09 \\
\hline
\end{tabular}
\tablefoot{The resulting rotational temperatures are consistently below the CMB level. $RV$ is measured in the barycentric rest frame.}
\end{table}

\section{Rotational model calculations}
\label{app:model_details}
\subsection{Selection rules}
For electronic transitions between a lower state X with the electronic orbit quantum number $\Lambda''$ and an excited state A with the electronic orbit quantum number $\Lambda'$, the electronic orbit quantum number 
$\Lambda$ can only change by $\Delta\Lambda$\,=\,$\Lambda' - \Lambda''$\,=\,$\pm 1$ or 0. 
In addition to the electrons, the nuclei in the molecule can rotate as well.
This rotation is characterised by the nuclear rotational quantum number $F$.
The total rotational quantum number is $J$\,=\,$\Lambda + F$.
This means that $J$ is always larger than or equal to $\Lambda$ in the respective state, so $J' \geq \Lambda'$ and $J'' \geq \Lambda''$.
The effect of those selection rules for the different electronic transitions is shown in Table \ref{tab:allowed_transitions}.

\subsection{Relative line strengths}
\begin{figure}[!ht]
\centering
\includegraphics[width=.93\hsize,trim={20 20 10 20},clip]{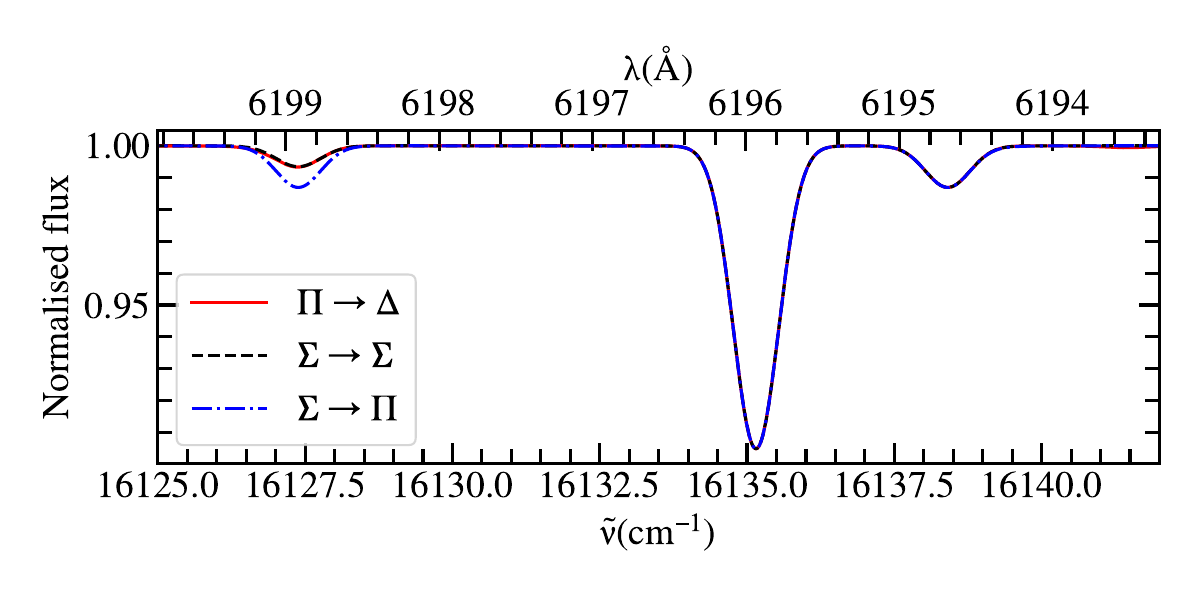}
\caption{Comparison between a $\Pi \rightarrow \Delta$ (red, $T_\mathrm{rot}=5.24\,$K), $\Sigma \rightarrow \Sigma$ (dashed black, $T_\mathrm{rot}=2.17\,$K) and $\Sigma \rightarrow \Pi$ (dash-dotted blue, $T_\mathrm{rot}=4.59\,$K) transition. 
The rotational temperatures are matched, so that the models have the same line ratio of the two strongest transitions. 
In this temperature range, the $\Pi \rightarrow \Delta$ and $\Sigma \rightarrow \Sigma$ models have the same line ratios, but different temperatures. The $\Sigma \rightarrow \Pi$ model has a stronger transition at lower wave numbers, which is as strong as the weak transition at higher wave numbers.}
\label{fig:model_comparison}
\end{figure}

The different electronic transitions can have different rotational temperatures or relative line strengths.
In Fig.~\ref{fig:model_comparison} we compare this effect for our used electronic transitions $\Sigma\rightarrow\Sigma$, $\Sigma\rightarrow\Pi$ and $\Pi\rightarrow\Delta$.
$\Sigma\rightarrow\Sigma$ and $\Pi\rightarrow\Delta$ show the same relative line strengths of the three strongest rotational transitions, but the $\Sigma\rightarrow\Sigma$ model needs a much lower $T_\mathrm{rot}$ to fit the same line ratios as the $\Pi\rightarrow\Delta$ model.
However, both have the common pattern of strong central absorption, weaker absorption at higher wave numbers, and even weaker absorption at lower wave numbers.
The $\Sigma\rightarrow\Pi$ is a different case, where the two weaker features R(1) and Q(1) are equally strong.
They have the same $J''$ and the same H\"onl-London factors of $A_{\Lambda''J''}$\,=\,1, which makes them equally strong for all $T_\mathrm{rot}$ (Fig.~\ref{fig:model_comparison}).

\subsection{Sensitivity to \texorpdfstring{$T_\mathrm{rot}$}\ }
\begin{figure}[!ht]
\centering
\includegraphics[width=.93\hsize,trim={20 20 10 20},clip]{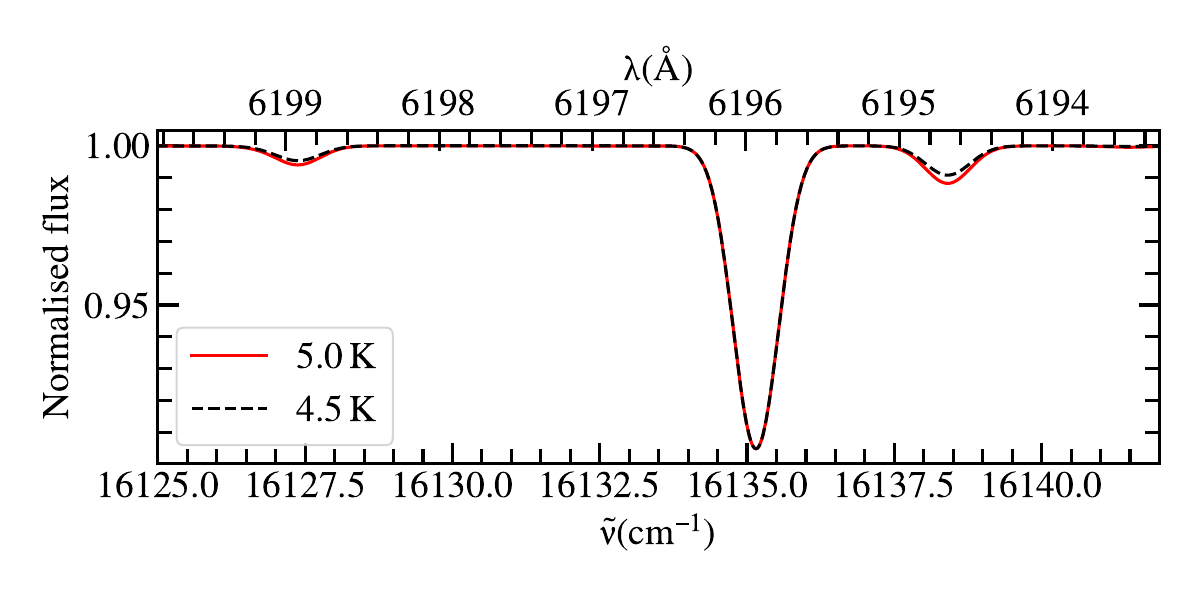}
\caption{$\Pi \rightarrow \Delta$ model at $T_\mathrm{rot}$\,=\,5.0\,K (red) and $T_\mathrm{rot}$\,=\,4.5\,K (dashed black), visualising the sensitivity of the model to changes in $T_\mathrm{rot}$.}
\label{fig:temp_comparison}
\end{figure}

To show the sensitivity of the models with respect to $T_\mathrm{rot}$, we show a comparison of different rotational temperatures in Fig.~\ref{fig:temp_comparison}.
A variation by 0.5\,K is roughly equal to 3$\sigma$ of the fitting errors in Table \ref{tab:fit_results_6196} and causes a change in the model spectrum, which is significant with respect to the S/N of the data. It should be also noted, that we fit the whole spectrum, so we should be able to reach higher sensitivity. 

\begin{table}[!h]
    \caption{Allowed rotational transitions for the respective electronic transitions.}
    \centering
    \begin{tabular}{cccc}
    \hline\hline
        Electronic transition & $J''_\mathrm{min}$ & $J'_\mathrm{min}$ & Min. $J$ rot. transitions\\
        \hline
        $\Sigma\rightarrow\Sigma$ & 0 & 0 & R(0), P(1)\\
        $\Sigma\rightarrow\Pi$ & 0 & 1 & R(0), Q(1), P(2)\\
        $\Pi\rightarrow\Delta$ & 1 & 2 & R(1), Q(2), P(3)\\
        \hline
    \end{tabular}
    \label{tab:allowed_transitions}
\end{table}

\section{Additional figures}
\label{app:add_figures}
%\setcounter{section}{2}
% \label{app:a}
\setcounter{figure}{0}

\begin{figure}[ht]
\centering
\subfloat[]{
\includegraphics[width=0.7\hsize,trim={20 20 10 20}]{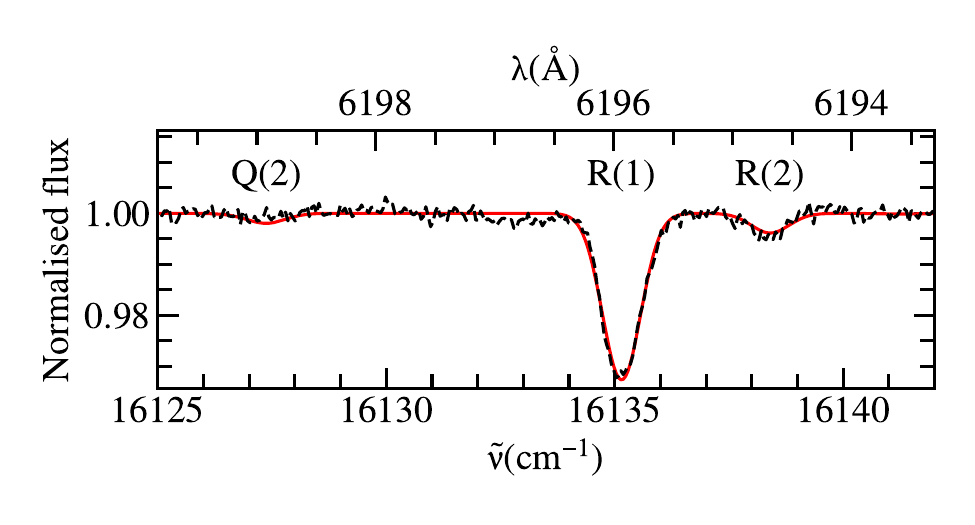}
     \label{fig:HD23180_6196}
}\\
\subfloat[]{
\includegraphics[width=0.7\hsize,trim={20 20 10 20}]{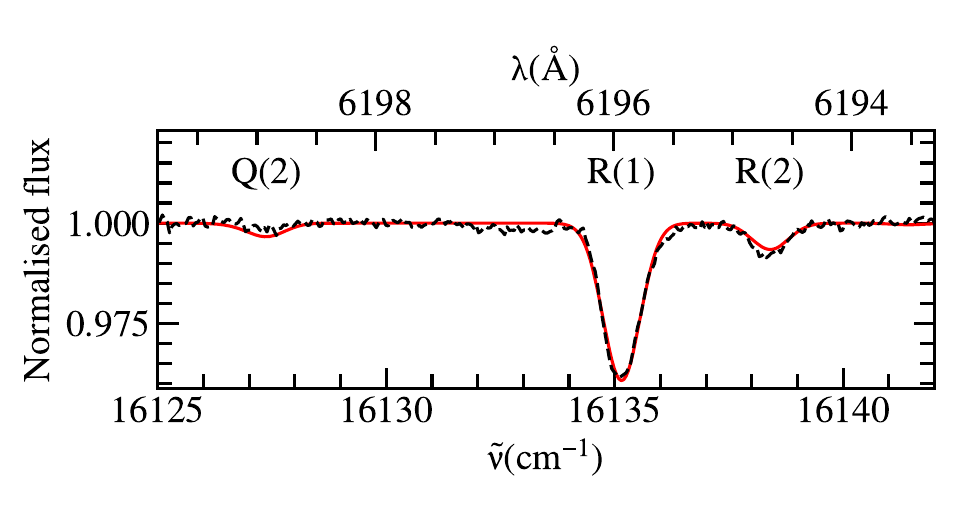}
     \label{fig:HD24398_6196}
}\\
\subfloat[]{
\includegraphics[width=0.7\hsize,trim={20 20 10 20}]{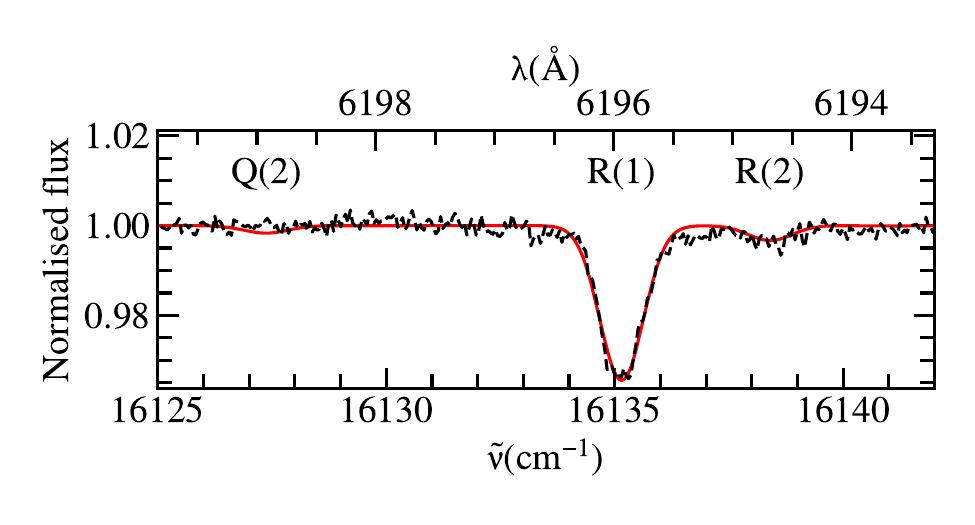}
     \label{fig:HD144470_6196}
}\\
\subfloat[]{
\includegraphics[width=0.7\hsize,trim={20 20 10 20}]{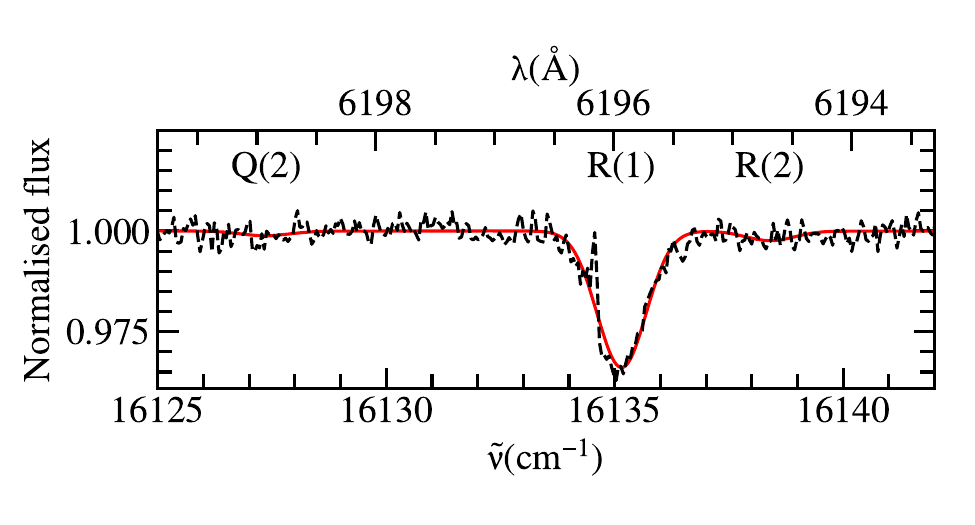}
     \label{fig:HD147165_6196}
}\\
\subfloat[]{
\includegraphics[width=0.7\hsize,trim={20 20 10 20}]{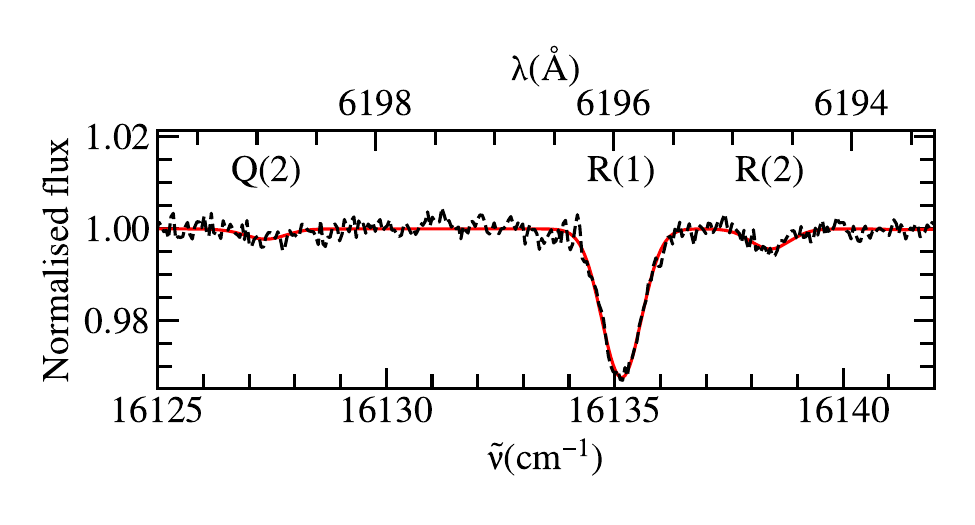}
     \label{fig:HD147683_6196}
}
\caption{Fit of the HD~23180 (a), HD~24398 (b), HD~144470 (c), HD~147165 (d), and HD~147683 (e) sight lines using a $\Pi\rightarrow\Delta$   transition of a linear molecule model for the 6196\,{\AA} DIB system.}
\label{fig:6196_fits_1}
\end{figure}

\begin{figure}[ht]
\centering
\subfloat[]{
\includegraphics[width=0.7\hsize,trim={20 20 10 20}]{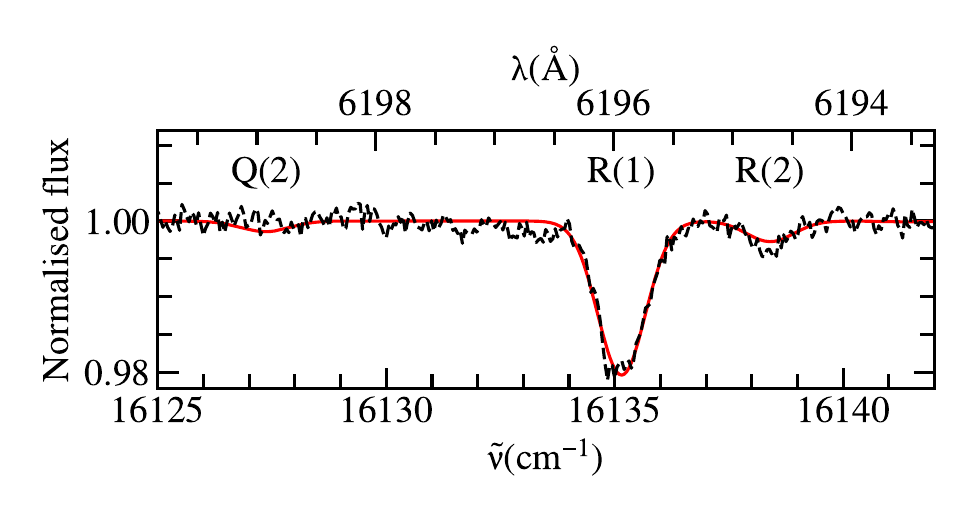}
     \label{fig:HD149757_6196}
}\\
\subfloat[]{
\includegraphics[width=0.7\hsize,trim={20 20 10 20}]{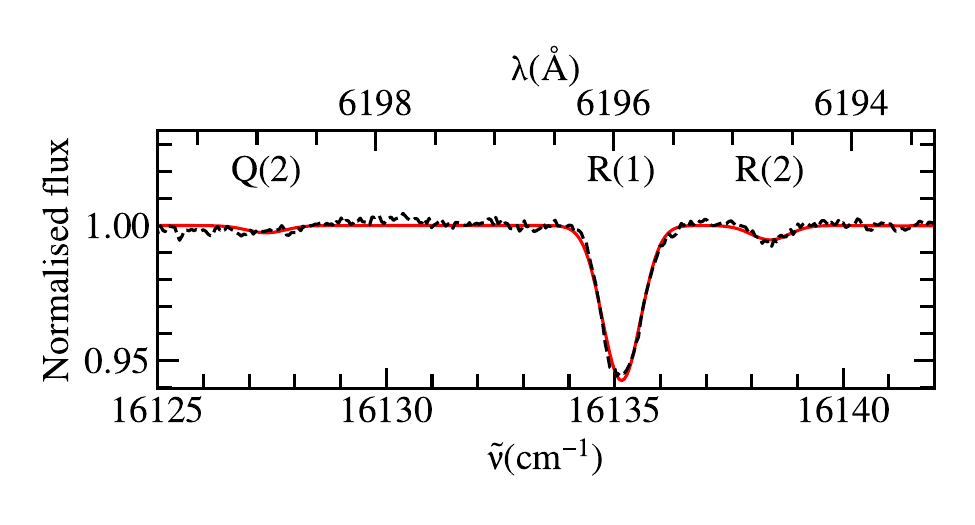}
     \label{fig:HD166937_6196}
}\\
\subfloat[]{
\includegraphics[width=0.7\hsize,trim={20 20 10 20}]{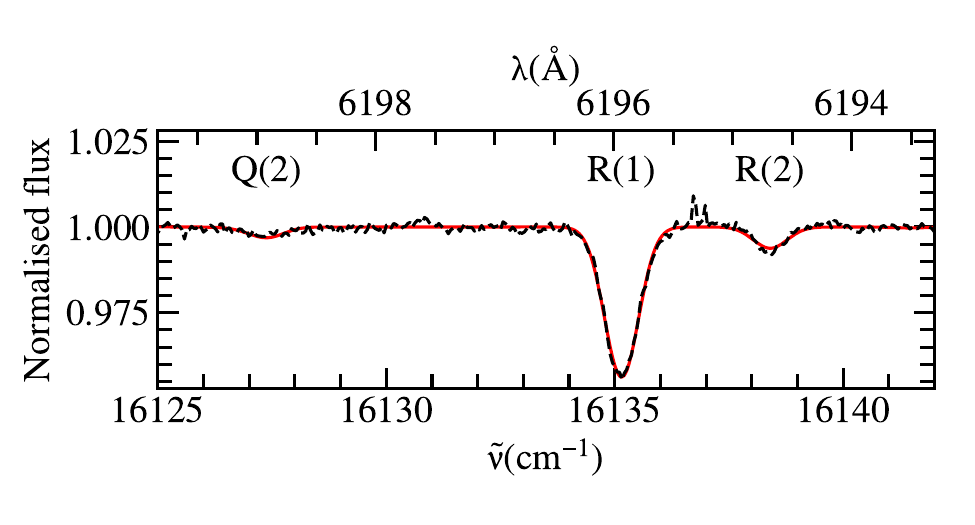}
     \label{fig:HD184915_6196}
}\\
\subfloat[]{
\includegraphics[width=0.7\hsize,trim={20 20 10 20}]{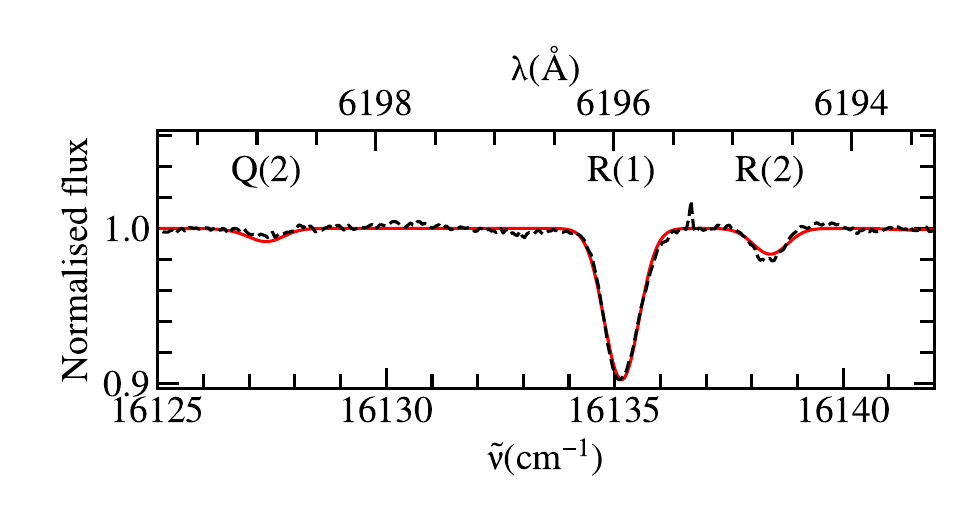}
     \label{fig:HD185418_6196}
}\\
\subfloat[]{
\includegraphics[width=0.7\hsize,trim={20 20 10 20}]{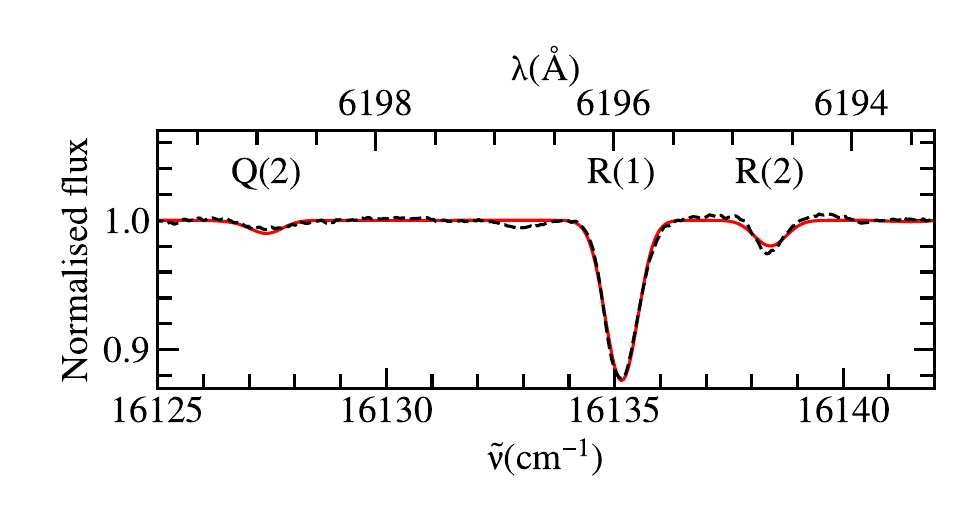}
     \label{fig:HD185859_6196}
}\\
\subfloat[]{
\includegraphics[width=0.7\hsize,trim={20 20 10 20}]{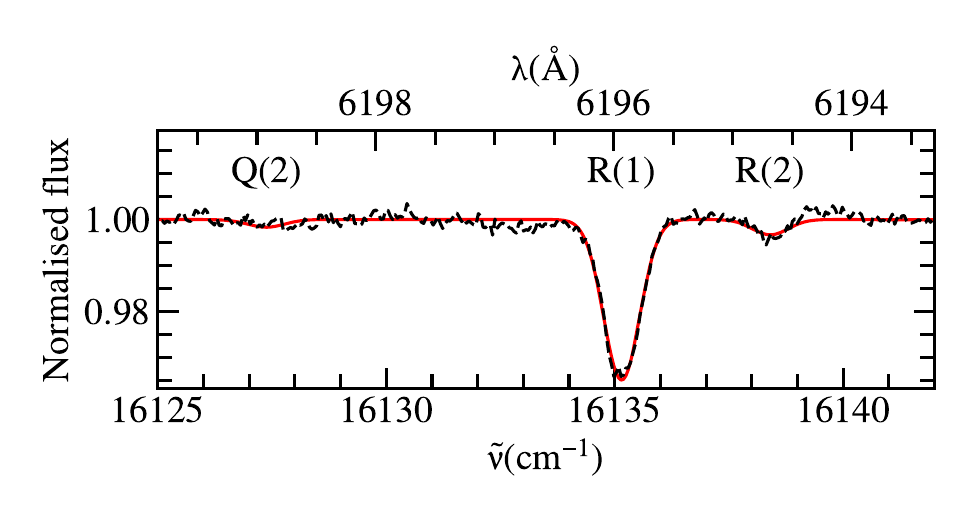}
     \label{fig:HD203532_6196}
}
\caption{Same as Fig.~\ref{fig:6196_fits_1}, but for the sight lines towards HD~149757 (a), HD~166937 (b), HD~184915 (c), HD~185418 (d), HD~185859 (e), and HD~203532 (f).}
\label{fig:6196_fits_2}
\end{figure}

\begin{figure}[ht]
\centering
\subfloat[]{
\includegraphics[width=0.7\hsize,trim={20 20 10 20}]{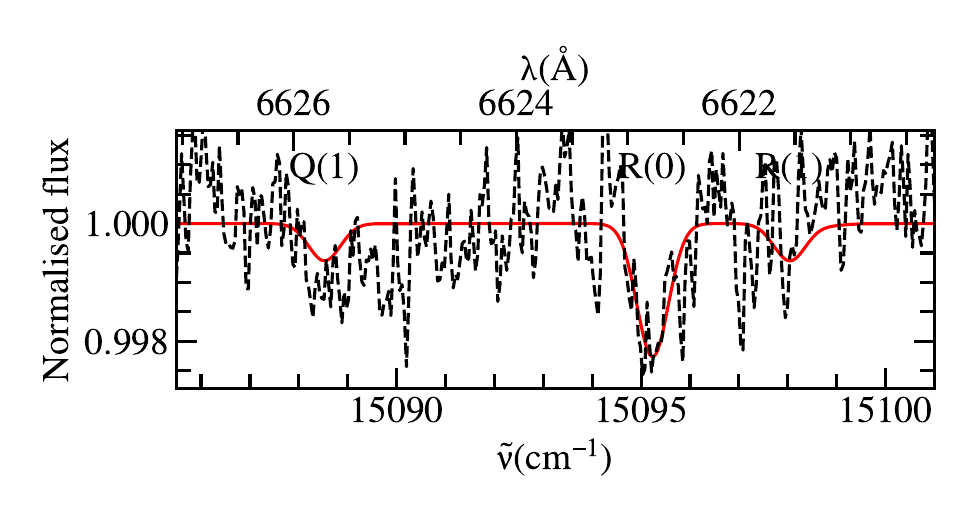}
     \label{fig:HD23180_6623}
}\\
\subfloat[]{
\includegraphics[width=0.7\hsize,trim={20 20 10 20}]{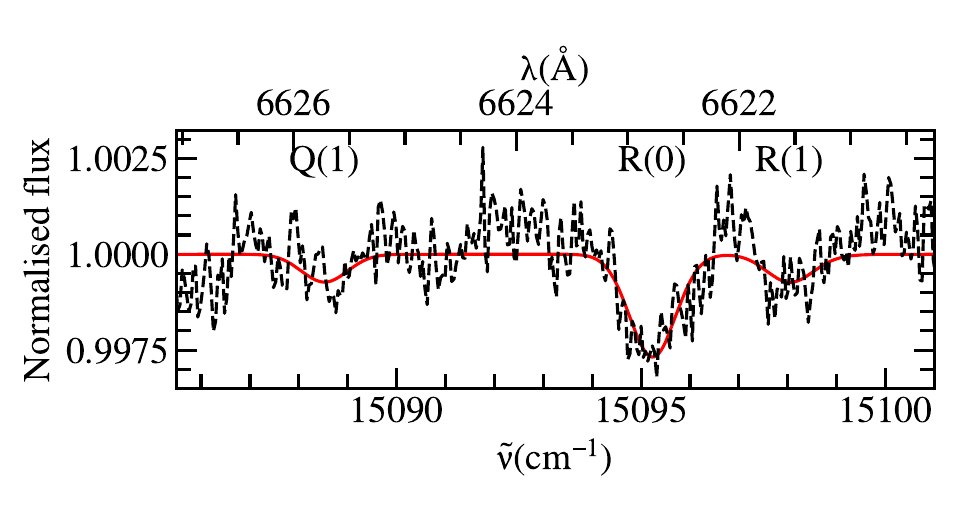}
     \label{fig:HD24398_6623}
}\\
\subfloat[]{
\includegraphics[width=0.7\hsize,trim={20 20 10 20}]{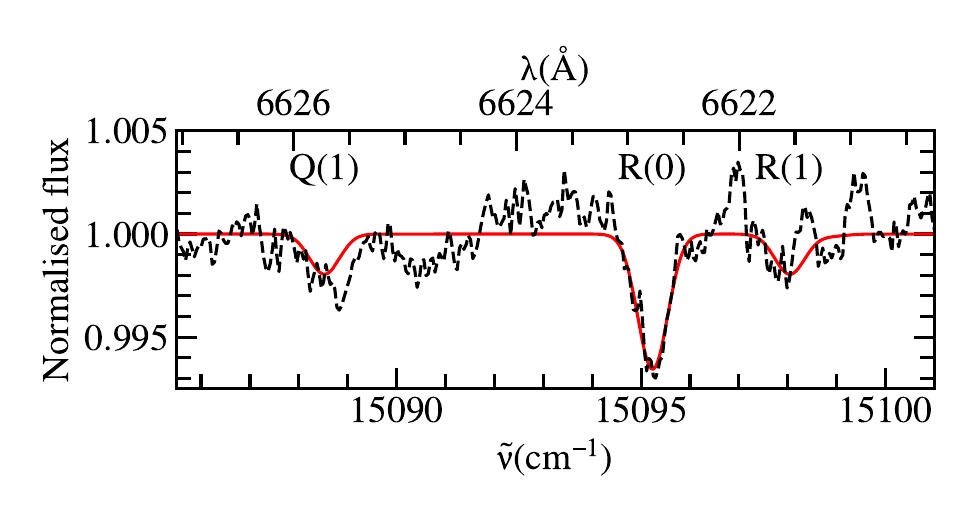}
     \label{fig:HD184915_6623}
}\\
\subfloat[]{
\includegraphics[width=0.7\hsize,trim={20 20 10 20}]{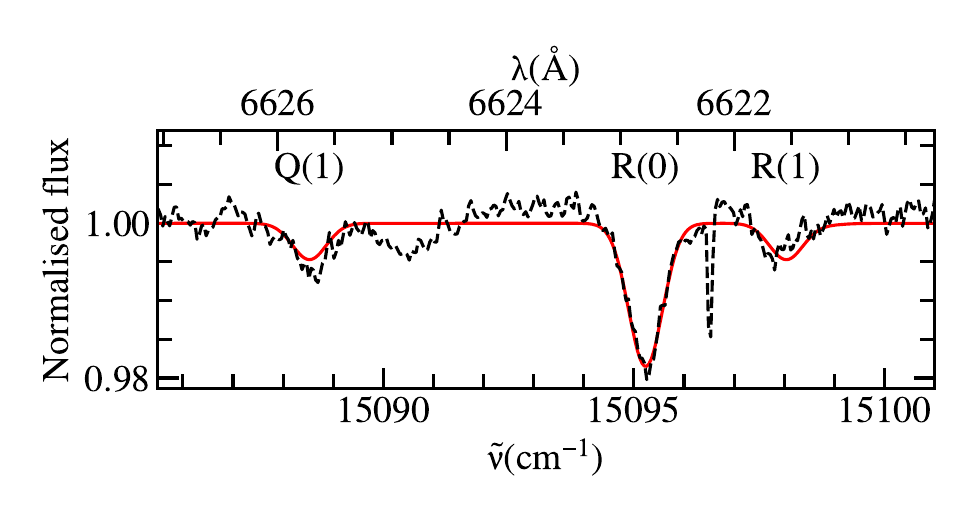}
     \label{fig:HD185418_6623}
}\\
\subfloat[]{
\includegraphics[width=0.7\hsize,trim={20 20 10 20}]{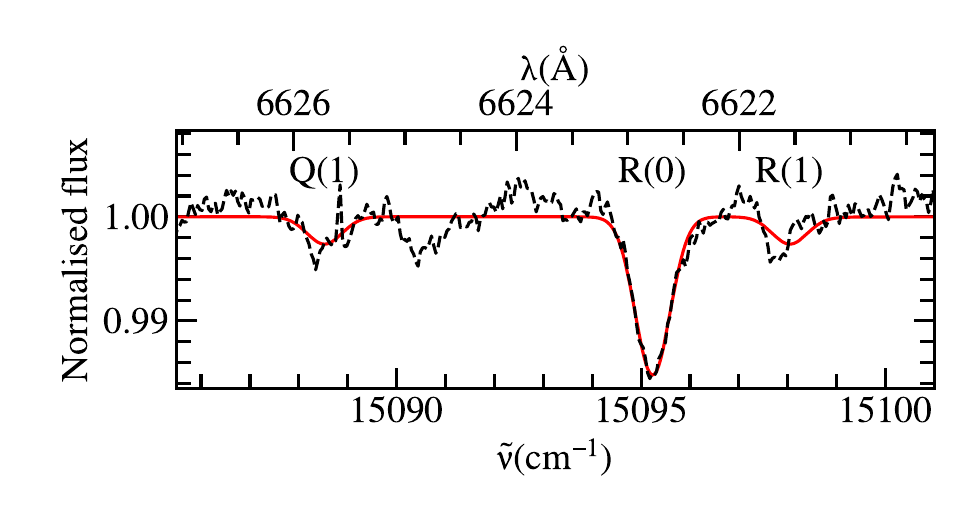}
     \label{fig:HD185859_6623}
}\\
\subfloat[]{
\includegraphics[width=0.7\hsize,trim={20 20 10 20}]{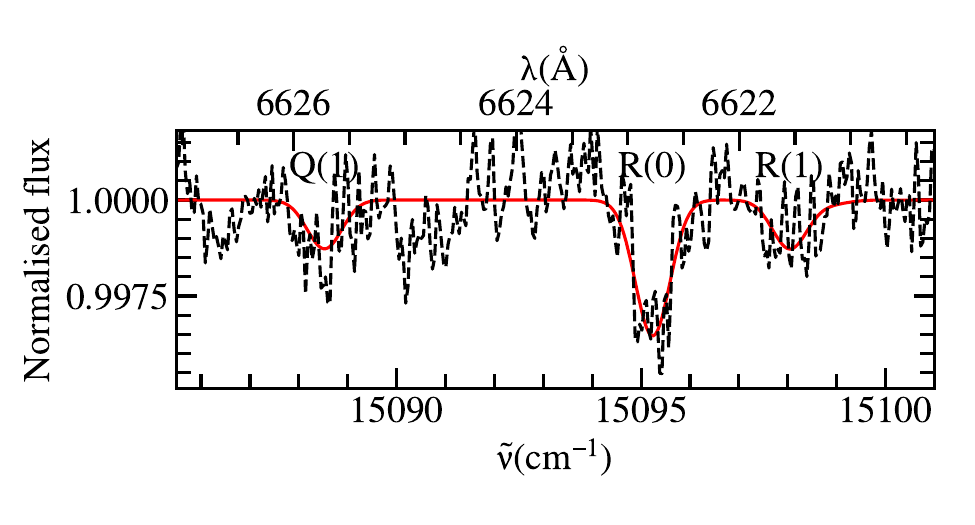}
     \label{fig:HD203532_6623}
}
\caption{Fit of the HD~23180 (a), HD~24398 (b), HD~184915 (c), HD~185418 (d), HD~185859 (e), and HD~203532 (f) sight lines using a $\Sigma\rightarrow\Pi$  transition of a linear molecule model for the 6623\,{\AA} DIB system.}
\label{fig:6623_fits}
\end{figure}

\begin{figure}[ht]
\centering
\includegraphics[width=0.9\hsize]{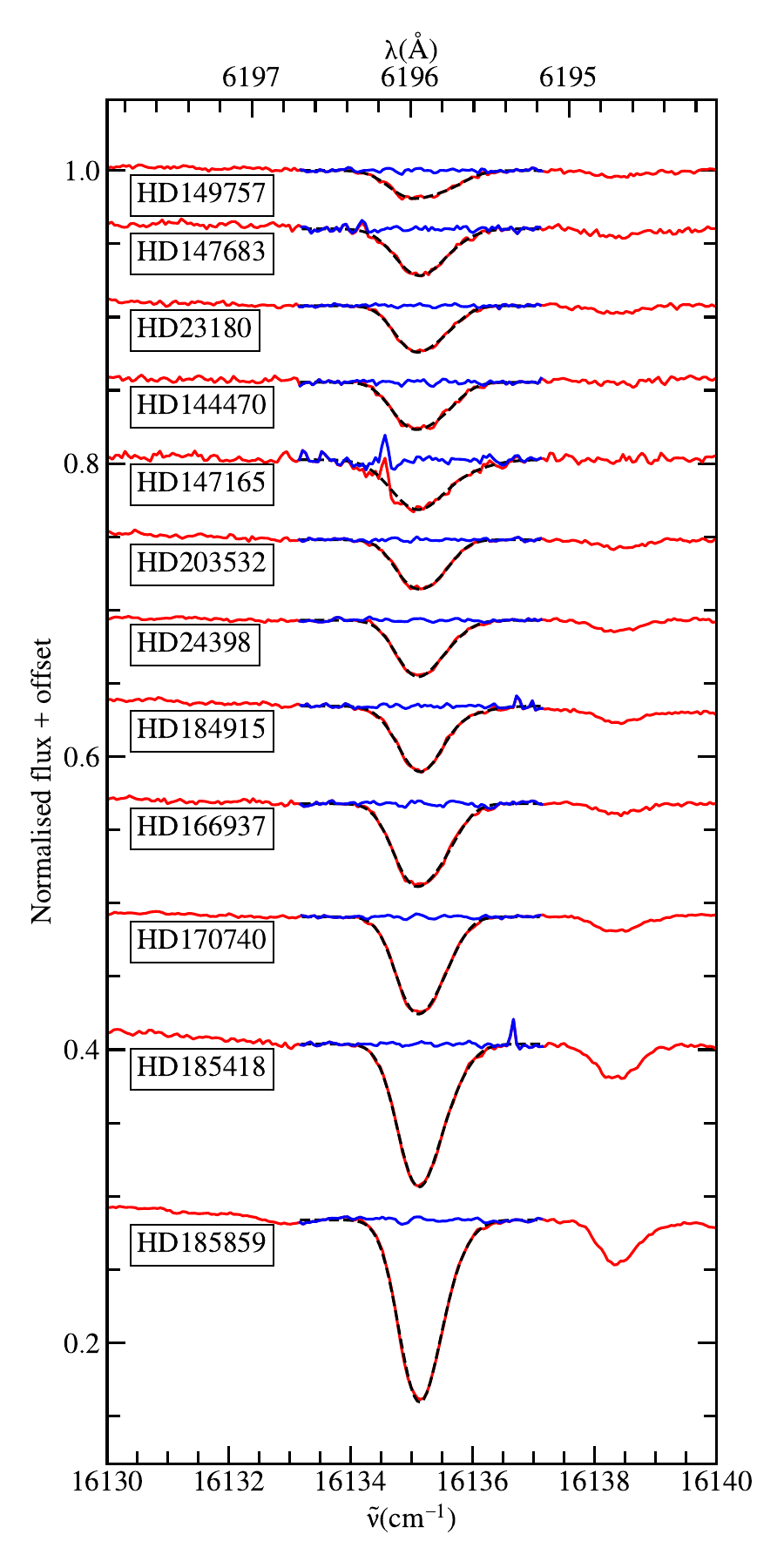}
\caption{Double-Gaussian fit (dashed black) of the 6196\,{\AA}~DIB in the spectra of our single cloud sight lines (red). The residual flux (blue) does not show significant systemic substructures of the 6196\,{\AA}~DIB.}
\label{fig:double_gauss_fits}
\end{figure}

\clearpage

\section{Deuterated DIB carrier candidates}
\label{app:deuterated}
A large number of candidates contains a deuterium atom, as can be seen in Table \ref{tab:carrier_candidates_deu}.
Deuterium might seem very unlikely due to its low abundance, but if some molecular formation channel is more efficient for deuterium than hydrogen, it cannot be excluded.
With an abundance ratio of [D]/[H] $\approx 10^{-5}$ \citep{1973ApJ...184L.101B}, deuterium is still abundant compared to species other than hydrogen and helium. 
Deuterated molecules compared with hydrogenated molecules have lower zero-point energies due to the molecule’s reduced mass being approximately doubled. This makes the deuterated molecule slightly more thermodynamically stable and is the underlying principle of fractionation in astrochemistry. And in cases such as N$_2$ \citep{refId0}, the fractionation can be photo-induced due to the slight differences in the photodissociation pathways that are slightly shifted by the change in zero-point energies in the heavier isotopologues.
Especially for the anionic species, no experimental data was available from literature, but some neutrals could be ruled out as carriers.
%SrD
In a rotational analysis of the A-X band system of SrD, \citet{1984PhyS...29..456A} experimentally determined a rotational constant of $B''$\,=\,1.86044(12)\,cm$^{-1}$ which is too far from our value of $B''$\,=\,2.05$\pm$0.07\,cm$^{-1}$ for a $\Sigma\rightarrow\Sigma$ transition.
%CaD
For CaD, both $B''$ and $B'$ are too large, and no transition is exactly at the observed energy, as measured experimentally by \citet{2012JMoSp.281...47G}.
%MgD and GaD CrD
The experimentally measured rotational constants of MgD of $B''$\,=\,3.0009462(11)\,cm$^{-1}$ \citep{1988JChPh..89..673L}, GaD of $B''$\,=\,3.085\,cm$^{-1}$ \citep{LAKSHMINARAYANA1987417} and CrD of $B''$\,=\,3.175443(21)\,cm$^{-1}$ \citep{1995JMoSp.172...91R} are too far from our observed values of the $\Sigma\rightarrow\Pi$ transitions with $B''$\,=\,3.36$\pm$0.09\,cm$^{-1}$ (6623~{\AA}~DIB system) and $B''$\,=\,3.47$\pm$0.12\,cm$^{-1}$ (6440~{\AA}~DIB system).
Additionally, no matching transition wavelengths were found for those species.
%ZnD
For ZnD, \citet{2017JMoSp.339...17G} found a matching rotational constant of $B''$\,=\,3.35035973(26)\,cm$^{-1}$, but the lowest excited electronic state A$^2\Pi$ is at $T_\nu$\,=\,23387.6200(4)\,cm$^{-1}$ which is too high.

\begin{table}[ht]
    \caption{Rotational constants of proposed deuterated carrier molecules, same as Table \ref{tab:carrier_candidates}.}
    \centering
    \begin{tabular}{lccl}
    \hline\hline
    Molecule & $B''$ & Molecular & Lit.\\
    & (cm$^{-1}$) & ground term &  \\
    \hline
    SrD$^-$ & 1.66 & $^1\Sigma^+$ & ...\\
    KD & 1.74 & $^1\Sigma^+$ & ... \\
    SrD & 1.85 & $^2\Sigma^+$ & f\tablefootmark{a}\\
    CaD$^-$ & 1.97 & $^1\Sigma^+$ & ...\\
    YD$^-$ & 2.12 & $^2\Sigma^+/^2\Pi$ & ...\\
    CaD & 2.18 & $^2\Sigma^+$ & f\tablefootmark{b}\\
    NaD$^-$ & 2.26 & $^2\Sigma^+$ & ...\\
    ZrD$^-$ & 2.27 & $^3\Pi$ & ...\\
    YD & 2.32 & $^1\Sigma^+$ & ...\\
    \hline
    MoD & 2.91 & $^6\Sigma^+$ & ...\\
    TiD$^{+}$ & 3.01 & $^3\Pi$/$^3\Sigma^-$ & ...\\
    CrD$^{-}$ & 3.01 & $^5\Sigma^+$ & ... \\
    MgD & 3.06 & $^2\Sigma^+$ & f\tablefootmark{c}\\
    GaD & 3.17 & $^1\Sigma^+$ & f\tablefootmark{d}\\
    CrD & 3.26 & $^6\Sigma^+$ & f\tablefootmark{e}\\
    (RhD$^{+}$) & 3.40 & $^4\Sigma^-$/$^4\Pi$ & ...\\
    RuD & 3.43 & $^4\Sigma^-$ & ...\\
    RuD$^{-}$ & 3.46 & $^3\Sigma^-$ & ...\\
    ZnD & 3.49 & $^2\Sigma^+$ & f\tablefootmark{f}\\
    FeD$^{+}$ & 3.61 & $^5\Sigma^+$ & ...\\
    (RuD$^{+}$) & 3.66 & $^3\Sigma^-$/$^3\Pi$ & ...\\
    CuD$^{-}$ & 3.70 & $^2\Sigma^+$ & ...\\
    CoD$^{+}$ & 3.81 & $^4\Pi$/$^4\Sigma^-$ & ...\\
    PdD$^{-}$ & 3.91 & $^1\Sigma^+$ & ...\\
    PdD$^{+}$ & 3.97 & $^1\Sigma^+$ & ...\\
    \hline
    \end{tabular}
    \label{tab:carrier_candidates_deu}
    \tablebib{
    \tablefoottext{a} {\citet{1984PhyS...29..456A}}; 
    \tablefoottext{b} {\citet{2012JMoSp.281...47G}};
    \tablefoottext{c} {\citet{1988JChPh..89..673L}};
    \tablefoottext{d} {\citet{LAKSHMINARAYANA1987417}};
    \tablefoottext{e} {\citet{1995JMoSp.172...91R}};
    \tablefoottext{f} {\citet{2017JMoSp.339...17G}}
    }
\end{table}

\clearpage

\end{appendix}
\end{document}